\documentclass[10pt,pre,aps,showpacs,twocolumn,preprintnumbers,amsmath,amssymb,superscriptaddress,longbibliography]{revtex4-1}
\usepackage{graphicx}    
\usepackage{latexsym}    %
\usepackage{color}       %
\usepackage{dcolumn}     
\usepackage{bm}          
\usepackage{amssymb}     
\usepackage{hyperref}    
\expandafter\let\csname equation*\endcsname\relax
\expandafter\let\csname endequation*\endcsname\relax
\usepackage{amsmath}
\usepackage{ulem}
\usepackage{mathtools}          

\usepackage{graphics}
\usepackage{epsfig}
\usepackage{subfigure}
\usepackage{nicefrac}
\usepackage{verbatim}
\usepackage{enumerate}
\usepackage{footnote}
\usepackage{booktabs}
\usepackage{siunitx}
\usepackage{xcolor}
\usepackage{ulem}
 \usepackage{soul}

\begin{document}
\title{Nonlocal Contributions to the Turbulent Cascade in Magnetohydrodynamic Plasmas}
\author{J.~Friedrich}
\affiliation{ForWind, Institute of Physics, University of Oldenburg,
Küpkersweg 70, D-26129 Oldenburg, Germany}
\author{M.~Wilbert}
\affiliation{Institut für Theoretische Physik I, Ruhr-Universität Bochum, Universitätsstraße 150, 44780 Bochum, Germany}
\author{R.~Marino}
\affiliation{CNRS, \'Ecole Centrale de Lyon, INSA de Lyon, Universit\'e Claude Bernard Lyon 1, Laboratoire de M\'ecanique des Fluides et d’Acoustique - UMR 5509, F-69134 Ecully, France}
\begin{abstract}
  We present evidence for nonlocal contributions to the turbulent energy cascade in magnetohydrodynamic (MHD) plasmas. Therefore, we revisit a well-known result derived directly from the MHD equations, i.e., the Politano \& Pouquet (P\&P) law for the transfer of kinetic and magnetic energy in scale. We propose adding a term that accounts for nonlocal transfer and represents the influence of fluctuations from large scales due to the Alfv\'en effect. Supported by direct numerical simulations of homogeneous and isotropic MHD turbulence, we verify that in some plasma configurations, neglecting the additional nonlocal term might consistently overestimate energy dissipation rates and, thus, the contributions of turbulent energy dissipation potentially affecting solar wind heating; a central puzzle in space plasma physics that motivates the present work.
\end{abstract}
\maketitle
\section{Introduction}
\label{sec:intro}
The physical mechanisms underlying the solar wind - a continuous flow of charged particles emitted from the solar corona - have been hypothesized since the beginning of the past century. Based on Biermann's observations of comet tail motion relative to the Sun~\cite{biermann1951kometenschweife}, Parker provided the first comprehensive theoretical description of the solar wind as a rapidly expanding outer coronal atmosphere overcoming the Sun's gravitational field~\cite{parker1958dynamics}. Assuming an adiabatic solar wind expansion, Parker also considered the radial temperature profile $T(r)$ as a function of the heliocentric distance $r$. Nonetheless, a purely adiabatic expansion of the solar wind, which suggests a radial temperature profile of proton temperature as $T(r) \sim r^{-4/3}$, differs from actual solar wind plasma measurements that exhibit a much slower decay with increasing distances $r$~\cite{Matthaeus1999,richardson2003radial,hundhausen2012coronal}.

A possible mechanism that would supply heat to the expanding solar wind - and thus explain the slower decay of the temperature profile - is the dissipation of turbulent kinetic and magnetic energy into heat predicted by a phenomenological description of magnetohydrodynamic (MHD) turbulence~\cite{Marino_2008} (we also refer the reader to the recent review article on scaling laws in solar wind turbulence~\cite{MARINO2023}).
The concept of turbulent cascades, which has been put forth in the context of hydrodynamic turbulence by Kolmogorov~\cite{Kolmogorov1941}, Heisenberg~\cite{Heisenberg1948}, von Weizs\"acker~\cite{Weizsacker1948}, and Onsager~\cite{Onsager1949}, highlights the fact that turbulent motions are essentially transport processes of energy in scale. The existence of energy cascades in fluid and plasma turbulence has been assessed experimentally (e.g., in channel flows and fusion plasmas~\cite{laboratory1,laboratory2}), as well as by atmospheric~\cite{atmosphere}, oceanic~\cite{ocean}, and space plasma measurements~\cite{space}. A central quantity in phenomenological descriptions of such cascade processes is the averaged energy dissipation rate $\langle \varepsilon_{kin} \rangle$, which is assumed to entirely characterize the transfer of energy from the injection scale, where the turbulent flow is stirred, to small scales where energy is dissipated into heat. 

\begin{figure}
    \centering
    \includegraphics[width=0.495 \textwidth]{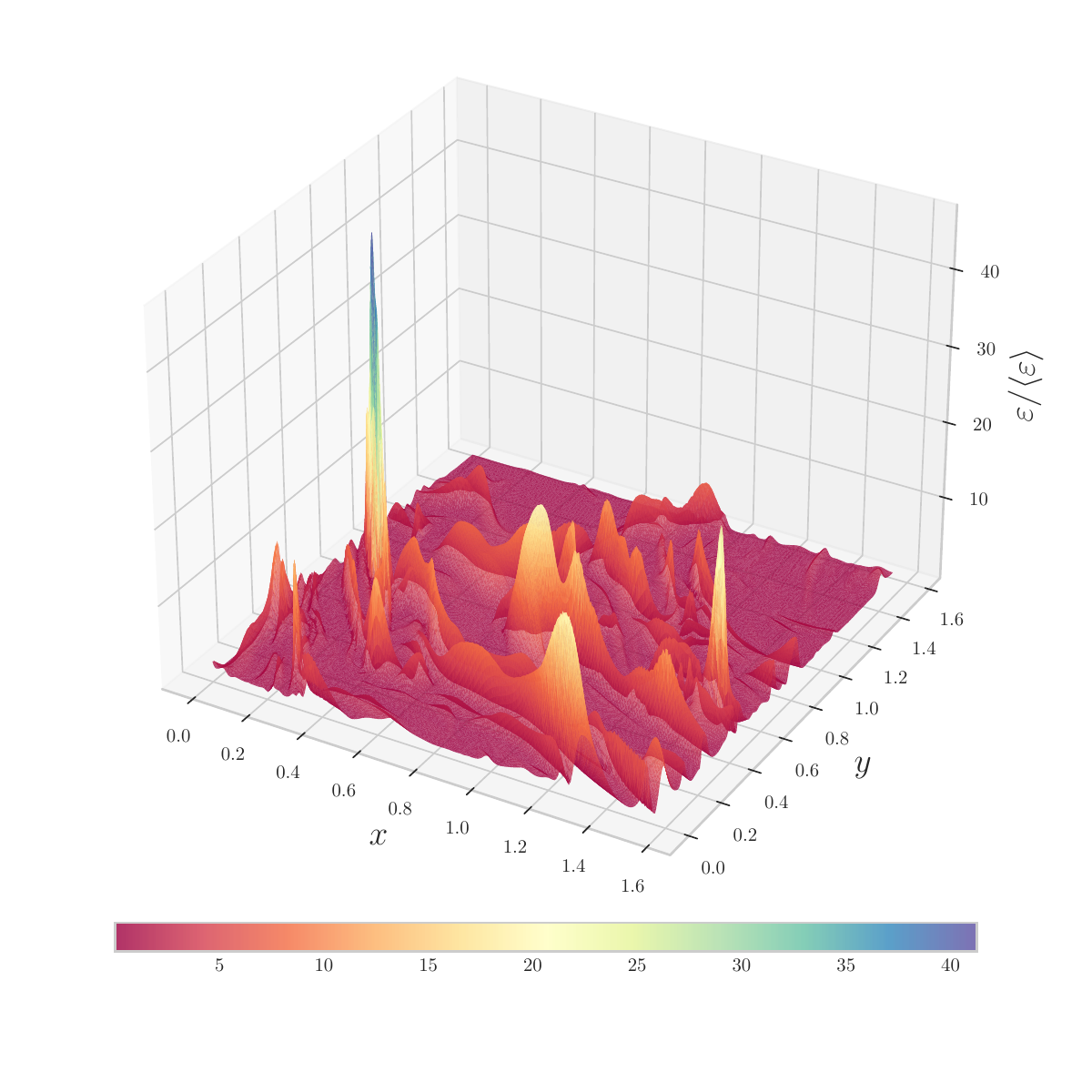}
    \includegraphics[width=0.495 \textwidth]{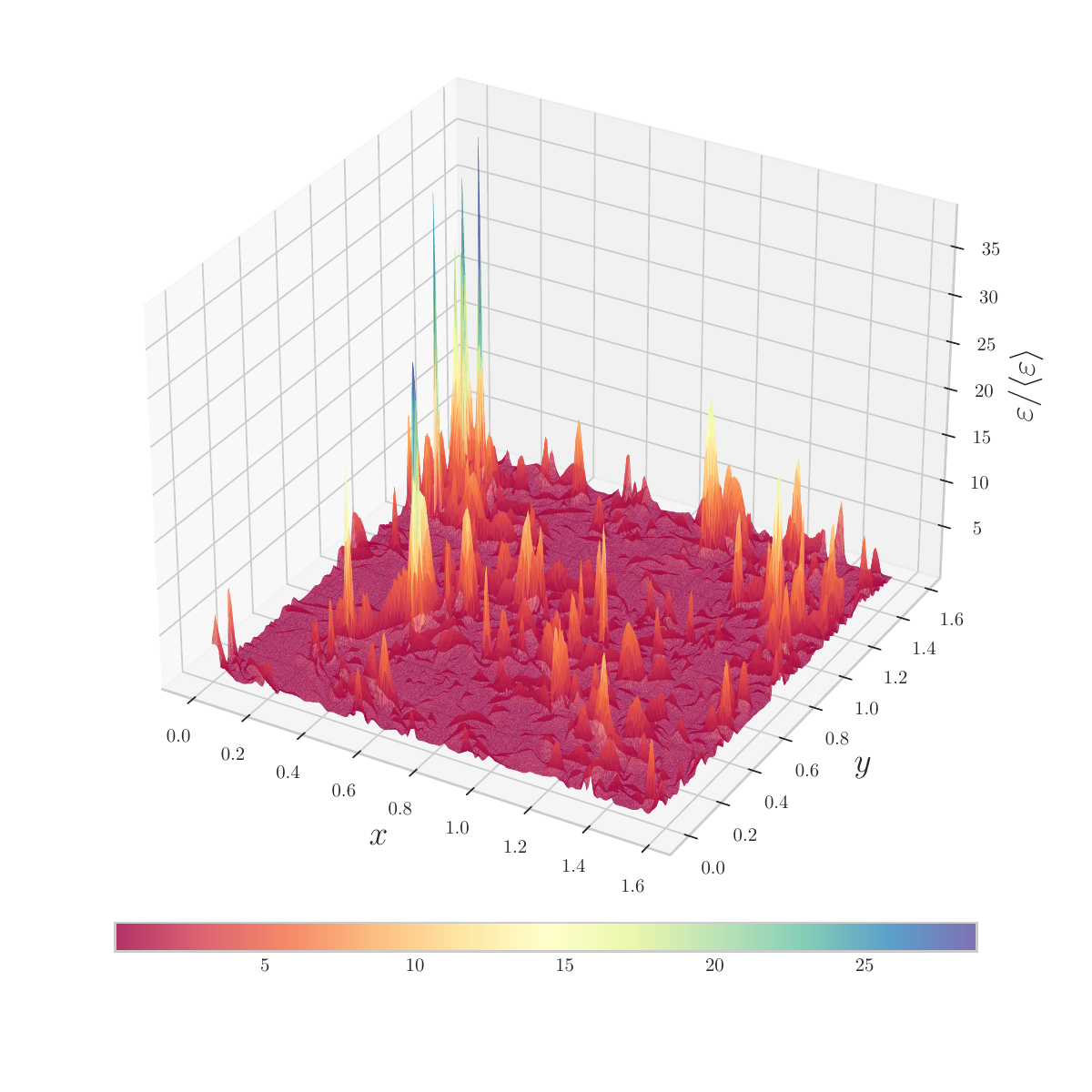}
     \caption{\emph{Top:} Surface plot of a cut through the total local energy dissipation rate, i.e., the sum of Eqs. (\ref{eq:eps_kin}) and (\ref{eq:eps_mag}), obtained from a direct numerical simulation of homogeneous and isotropic MHD turbulence with $1024^3$ spatial points  (here only an area of $256 \times 256$ points is shown) with $\textrm{Re}_{\lambda, kin}$ and $\textrm{Re}_{\lambda, mag}$ and  (see main text and Tab.~\ref{tab:charac} for additional information). The organization of the dissipation field into sheet-like structures is visible and is dominated by a singular sheet peaking at roughly 40 times the mean value. \emph{Bottom:} Same plot as on top, but obtained from a snapshot of a direct numerical simulation of homogeneous and isotropic hydrodynamic turbulence provided by the Johns Hopkins turbulence database JHTDB (\textsc{http://turbulence.pha.jhu.edu}) with the same spatial resolution $1024^3$ and $\textrm{Re}_{\lambda, kin}=418$ (again, only an area of $256 \times 256$ points is shown). The dissipation field is organized more randomly than its MHD counterpart.}
    \label{fig:diss_rates}
\end{figure}

In a conducting fluid, the rate at which kinetic and magnetic energy is dissipated is determined by the local kinetic and magnetic energy dissipation rates
\begin{align}
    \label{eq:eps_kin}
     \varepsilon_{kin}({\bf x},t) =& \frac{\nu}{2} \sum_{i,k} \left(\frac{\partial u_i({\bf x},t)}{\partial x_k} +\frac{\partial u_k({\bf x},t)}{\partial x_i} \right)^2\;,\\
     \varepsilon_{mag}({\bf x},t) =& \frac{\lambda}{2} \sum_{i,k} \left(\frac{\partial h_i({\bf x},t)}{\partial x_k} +\frac{\partial h_k({\bf x},t)}{\partial x_i} \right)^2\;,
     \label{eq:eps_mag}
\end{align}
where ${\bf u}({\bf x},t)$ denotes the velocity field,  ${\bf h}({\bf x},t)=\sqrt{\mu / 4\pi \rho}\;  {\bf H}({\bf x},t)$ the rescaled magnetic field, $\rho$ the density of the fluid, $\mu$ the permeability, $\nu$ the kinematic viscosity, and $\lambda$ the magnetic diffusivity (see Appendix~\ref{app:energy_balance} for further details). Fig.~\ref{fig:diss_rates} (a) depicts a two-dimensional cut through the total energy dissipation rate field $\varepsilon_{tot}({\bf x},t) =\varepsilon_{kin}({\bf x},t) +\varepsilon_{mag}({\bf x},t)$ from a direct numerical simulation (DNS) of MHD turbulence (see Tab.~\ref{tab:charac} for further details). The dissipation field is highly fluctuating and exhibits strong gradients organized in current sheets in contrast to hydrodynamic turbulence \cite{marino_22}; see Fig.~\ref{fig:diss_rates} (b), where the peaks are more localized, hinting at the presence of vortex tubes~\cite{ishihara-gotoh-etal:2009,tanaka1993characterization}. 

The solar wind is weakly collisional, which implies that energy cascades from the MHD scales further down to proton and electron scales~\cite{matthaeus2020pathways}. Due to the limited resolution of the instruments onboard most spacecraft and the presence of measurement noise, strong gradients of the energy dissipation at kinetic scales can typically not be resolved directly from {\it in situ} solar wind observations. Nonetheless, the scale-wise energy transfer at the scales at which the interplanetary plasma has a fluid-like behavior can be estimated indirectly from third-order moments of velocity and magnetic fields \cite{Carbone_2009,MARINO2023}.
This can best be illustrated by considering the limit of vanishing magnetic field in the MHD equations, i.e., the Navier-Stokes equation. Under the assumption of homogeneity, isotropy, and statistical stationarity of the flow, the third-order moment of the longitudinal velocity increment $\delta_r u = [{\bf u}({\bf x}+{\bf r},t)-{\bf u}({\bf x},t)]\cdot \frac{{\bf r}}{r}$ is related to the averaged kinetic energy dissipation rate according to
\begin{equation}
S_{r\,r\,r}^{\mathbf{u}\mathbf{u} \mathbf{u}}(r)= \left \langle (\delta_r u)^3\right \rangle= -\frac{4}{5} \left \langle \varepsilon_{kin} \right \rangle r\;,
\label{eq:4/5_hydro}
\end{equation}
where angle brackets $\langle \ldots \rangle$ denote a suitable averaging procedure, e.g., ensemble averages~\cite{monin}. A similar law has been derived from the MHD equations based on the Elsässer fields by Politano and Pouquet~\cite{Politano1998,Politano1998b} and will hereafter be referred to as P\&P law. The P\&P law, which has been successfully used to assess the significance of solar wind heating~\cite{Marino_2008,Stawarz2010}, suggests that - similar to hydrodynamic turbulence - the nonlinear interactions during energy transfer are purely \emph{local} in scale $r$. The localness of interactions, however, in MHD turbulence is broken by the Alfv\'en effect~\cite{biskamp}, which implies that small-scale fluctuations are susceptible to large-scale magnetic field structures and behave approximately as Alfv\'en waves~\cite{alfven} - a central assumption in both phenomenological models put forth by Iroshnikov~\cite{iroshnikov1964turbulence} and Kraichnan~\cite{kraichnan1965inertial}. The Alfvénisation of small-scale fluctuations is also implied in the phenomenological model by Boldyrev~\cite{boldyrev:2005}, who suggests a scale-dependent alignment between velocity and magnetic field that effectively suppresses nonlinear transfer at smaller scales (we also refer to~\cite{alexakis2018cascades,molokov2007turbulence,nazarenko2011critical} for further discussions as well as the incorporation of potential anisotropies captured by the Goldreich-Sridhar model~\cite{goldreich1995toward}). The solar wind can be highly Alfv\'enic, supporting the propagation of outward and inward Alfv\'en waves \cite{Adhikari,Damicis_2015} (with respect to the Sun). At the same time, heliospheric plasmas develop a strong turbulent state leading to the apparent paradox known as Alfv\'enic turbulence~\cite{Veltri,Bavassano}. This condition, characterized by the simultaneous presence of strong turbulence and non-negligible correlations between velocity and the frozen-in magnetic field, is routinely observed in the solar wind~\cite{Bavassano_1998}. Alfv\'enicity, as well as homogeneity, isotropy, and compressibility, are space plasma features that vary throughout the heliosphere and in time, thus with heliocentric distance, heliolatitude, and the solar activity itself~\cite{Stawarz2010,Marino2011,Marino2012,Coburn_2012}. In other words, solar wind turbulence explores a vast parameter space of the magnetohydrodynamic cascade, driven by local nonlinear couplings \cite{aluie2010scale,mininni2011scale}. In specific conditions, for instance, due to the presence of large-scale shear and structures propagating from the Sun, cascade processes might be affected by interactions that are nonlocal in scale. Such effects should be reproduced by the third-order model we propose here - at least in the locally isotropic and homogeneous case. Beyond the MHD regime,  
turbulent heating in the solar wind results from different contributions, for instance, Landau-type damping~\cite{Cerri_damping} and non-resonant damping mechanisms (such as stochastic heating) occurring at small scales due to the existence of propagating plasma waves, but also the dissipation related to the presence of coherent structures originating from magnetic reconnection\cite{Cerri_reconnection} and currents~\citep{Vasquez2007}. Such kinetic effects, as well as anisotropy and compressibility, have been incorporated in recent modifications of the scaling laws used to assess solar wind heating~\cite{banerjee2013exact,galtier2005spectral}. Furthermore, strong turbulence dominated by nonlinear coherent structures such as Orszag-Tang vortices \citep{orszag_tang_1979,Foldes2023} and Alfvén vortices \citep{Shukla1985,Owen2016} imply the possibility of nonlocal couplings at MHD scales and thus contribute to energy transfer and dissipation in space plasmas. 

This article presents new evidence for nonlocal contributions to kinetic and magnetic energy transfer. To this end, we first revisit the derivation of the P\&P law~\cite{Politano1998} in Sec.~\ref{sec:derivation}, obtaining from first principles an additional nonlocal term that accounts for the influence of magnetic fields induced at large scales on fluctuations at small scales as suggested by the Iroshnikov-Kraichnan phenomenology. The importance of this nonlocal term for estimations of turbulent energy transfer based on third-order moments is then assessed by DNS of homogeneous and isotropic MHD turbulence in Sec.~\ref{sec:dns}, and the validity of the model proposed is demonstrated in the plasma under study. Although the original P\&P law has been verified in numerous DNS of two- and three-dimensional MHD turbulence~\cite{mininni2009finite,sorriso2002analysis,yousef2007exact} (see also~\cite{MARINO2023} for further references), here, we provide analytical and numerical evidence that there can be plasma configurations for which considering the nonlocal contributions to the MHD cascade is critical to correctly infer the plasma heating rate through the third-order law approach. These configurations can be representative of heliospheric plasmas, whose highly dynamical state is such that solar wind may at times be characterized by enhanced turbulence and nonlocal couplings of kinetic and magnetic modes, due to magnetic reconnection \cite{Cerri2}, magnetic switchbacks \cite{Hernandez_2021,Telloni_2022} or the interaction between fast and slow solar wind streams in the ecliptic \cite{Marino2011}.

\section{Revisiting the P\&P law using the invariant theory of MHD turbulence}
\label{sec:derivation}
In this section, we revisit the P\&P law~\cite{Politano1998,Politano1998b} using the invariant theory of MHD turbulence devised by Chandrasekhar~\cite{chandra:1951}.
Following standard statistical treatments of turbulent flows~\cite{monin}, we introduce ensemble averages denoted by the brackets $\langle \ldots \rangle$. Under the assumption of homogeneity, it is straightforward to derive an evolution equation for the total energy $E_{tot}(t)= \frac{1}{2}\langle u_i u_i + h_i h_i \rangle$ according to
\begin{equation}
    \dot E_{tot}(t)= \frac{3}{2}\frac{\textrm{d}}{\textrm{d}t}\left( u_{rms}^2 +h_{rms}^2 \right) =-\langle \varepsilon_{tot} \rangle\;,
    \label{eq:E_tot}
\end{equation}
where we imply summation over identical indices and introduce the root mean square velocity and magnetic fields (please see Appendix~\ref{app:energy_balance} for further details on the derivation of Eq. (\ref{eq:E_tot})).
In his seminal work, Chandrasekhar generalized the work of von K\'arm\'an and Howarth~\cite{DeKarman1938} to an invariant theory of homogeneous and isotropic MHD turbulence~\cite{chandra:1951}. 
Here, we consider evolution equations for the velocity and magnetic field correlation tensors, $C_{i\;j}^{\mathbf{u}\mathbf{u}}({\bf r},t)=\left \langle u_i u_j'  \right \rangle=\left \langle u_i({\bf x},t) u_j({\bf x}+{\bf r},t) \right \rangle$ and $C_{i\;j}^{\mathbf{h}\mathbf{h}}({\bf r},t)=\left \langle h_i h_j'  \right \rangle=\left \langle h_i({\bf x},t) h_j({\bf x}+{\bf r},t) \right \rangle$, which can be derived from the MHD equations (under the assumption of homogeneity, see Appendix~\ref{app:friedmann} and~\ref{app:kh}) according to
\begin{align}\label{eq:uu}
    \frac{\partial}{\partial t} \left \langle  u_i u_j'  \right \rangle -&  2\frac{\partial}{\partial r_k} \left \langle (u_i u_k-h_i h_k)  u_j' \right\rangle = 2 \nu \nabla_{{\bf r}}^2 \left \langle  u_i u_j'  \right \rangle\;,\\ 
     \frac{\partial}{\partial t} \left \langle  h_i h_j'  \right \rangle -&  2\frac{\partial}{\partial r_k} \left \langle (h_i u_k - h_k u_i ) h_j' \right \rangle = 2 \lambda \nabla_{{\bf r}}^2 \left \langle  h_i h_j'  \right \rangle\;.
     \label{eq:hh}
\end{align}
Here, tensors of third order in Eq. (\ref{eq:uu}), i.e.,  $C_{(ik)j}^{\mathbf{u}\mathbf{u}{\mathbf{u}}}({\bf r},t)= \left \langle u_i u_k u_j' \right \rangle$ and $C_{(ik)j}^{\mathbf{h}\mathbf{h}{\mathbf{u}}}({\bf r},t)= \left \langle h_i h_k u_j' \right \rangle$  are symmetric in the indices $i$ and $k$, which is represented by the round brackets $(ik)$. By contrast, the third order tensor in Eq. (\ref{eq:hh}) is anti-symmetric in $i$ and $k$, which is denoted by square brackets $[ik]$. It thus admits a different tensorial form than the latter two (see Appendix~\ref{app:chan} for further details), namely,
\begin{align}\nonumber
    A_{[ki]j}^{\mathbf{u}\mathbf{h}{\mathbf{h}}}({\bf r},t)=& \left \langle (h_i u_k - h_k u_i ) h_j' \right \rangle \\
    =& A_{[rt]t}^{\mathbf{u}\mathbf{h}{\mathbf{h}}}(r,t) \left(\frac{r_i}{r}\delta_{jk}- \frac{r_k}{r}\delta_{ij}\right)\;.
    \label{eq:antisym}
\end{align}
Furthermore, the indices $r$ and $t$ denote the longitudinal and transverse projections,
${\bf u}_r= \frac{{\bf r}}{r}\left(\frac{{\bf r}}{r} \cdot {\bf u} \right)$ and ${\bf u}_t= -\left(\frac{{\bf r}}{r} \times \left(\frac{{\bf r}}{r} \times {\bf u} \right)\right)$
respectively (the same projections hold for the magnetic field ${\bf h}({\bf x},t)$, see Appendix~\ref{app:chan} for further derivation).
Following Chandrasekhar~\cite{chandra:1951} (see Appendix~\ref{app:kh} for further derivations), we 
obtain evolution equations for the longitudinal velocity and magnetic field correlation functions 
\begin{widetext}
\begin{align}\label{eq:kh_u}
    \frac{\partial}{\partial t} C_{r\;r}^{\mathbf{u}\mathbf{u}}(r,t) =& \frac{1}{r^4}\left[\frac{\partial}{\partial r}r^4  \left( C_{r\;r\;r}^{\mathbf{u}\mathbf{u} \mathbf{u}}(r,t)-
    C_{r\;r\;r}^{\mathbf{h}\mathbf{h} \mathbf{u}}(r,t) + 2 \nu \frac{\partial}{\partial r} C_{r\;r}^{\mathbf{u}\mathbf{u}}(r,t)\right) \right]\;, \\
       \frac{\partial}{\partial t} C_{r\;r}^{\mathbf{h}\mathbf{h}}(r,t) =& -\frac{4}{r} A_{[rt]t}^{\mathbf{u}\mathbf{h} \mathbf{h}}(r,t)+ 2 \lambda \frac{1}{r^4} \frac{\partial}{\partial r} r^4 \frac{\partial}{\partial r} C_{r\;r}^{\mathbf{h}\mathbf{h}}(r,t) \;.
       \label{eq:kh_h}
\end{align}
\end{widetext}
Eq. (\ref{eq:kh_u}) is the generalization of the von K\'arm\'an-Howarth equation to MHD turbulence and the additional third-order correlation is due to the Lorentz force $C_{r\;r\;r}^{\mathbf{h}\mathbf{h} \mathbf{u}}(r,t)$ in the MHD equations. A similar equation also holds for the second-order longitudinal magnetic field correlation function (\ref{eq:kh_h}) where the third-order correlation is due to the advection by the velocity field and the corresponding transformation of the electric field ${\bf E}' ={\bf E}+ \mu {\bf u} \times {\bf H}$. Similar equations were derived recently in the context of Hall MHD~\cite{galtier2008karman}, where the Hall term modifies such a transformation.

To derive the equivalent of the four-fifths law (\ref{eq:4/5}) for MHD turbulence, we introduce the second and third-order longitudinal structure functions 
\begin{align}
S_{r\;r}^{{\bf u}{\bf u}}(r,t) =& \left \langle \left( \left[{\bf u}({\bf x}+{\bf r},t)- {\bf u}({\bf x},t) \right]\cdot \frac{{\bf r}}{r} \right)^2\right  \rangle\;, \\
S_{r\;r}^{{\bf h}{\bf h}}(r,t) =& \left \langle \left( \left[{\bf h}({\bf x}+{\bf r},t)- {\bf h}({\bf x},t) \right]\cdot \frac{{\bf r}}{r} \right)^2\right  \rangle\;, \\
    S_{r\;r\;r}^{{\bf u}{\bf u}{\bf u}}(r,t) =& \left \langle \left( \left[{\bf u}({\bf x}+{\bf r},t)- {\bf u}({\bf x},t) \right]\cdot \frac{{\bf r}}{r} \right)^3\right  \rangle\;.
\end{align}
As shown in Appendix~\ref{app:struc}, these structure functions are related to the longitudinal correlation functions as
\begin{align}
\label{eq:struc_corr}
S^{{\bf u}{\bf u}}_{r\;r}(r,t)=&2[ C_{r\;r}^{\mathbf{u}\mathbf{u}}(0,t)-C_{r\;r}^{\mathbf{u}\mathbf{u}}(r,t)]\;,\\
S^{{\bf h}{\bf h}}_{r\;r}(r,t)=&2[ C_{r\;r}^{\mathbf{h}\mathbf{h}}(0,t)-C_{r\;r}^{\mathbf{h}\mathbf{h}}(r,t)]\;,\\
S^{{\bf u}{\bf u}{\bf u}}_{r\;r\;r}(r,t)=&6 C_{r\;r\;r}^{\mathbf{u}\mathbf{u} \mathbf{u}}(r,t)\;.
\end{align}
Furthermore, we evaluate the longitudinal correlation functions for ${r}=0$, which yields
\begin{equation}
    C_{r\;r}^{\mathbf{u}\mathbf{u}}(0,t)= u_{rms}^2\;, \quad \textrm{and}\ \quad C_{r\;r}^{\mathbf{h}\mathbf{h}}(0,t)= h_{rms}^2\;.
    \label{eq:rel_rms}
\end{equation}
%
Combining Eqs. (\ref{eq:E_tot}), (\ref{eq:kh_u},  (\ref{eq:kh_h}), and (\ref{eq:struc_corr}-\ref{eq:rel_rms}) thus yields
\begin{widetext}
\begin{align} \nonumber
    \frac{1}{2}\frac{\partial}{\partial t} \left[S^{{\bf u}{\bf u}}_{r\;r}(r,t) + S^{{\bf h}{\bf h}}_{r\;r}(r,t) \right] =&-\frac{2}{3}
    \langle \varepsilon_{tot} \rangle-
    \frac{1}{r^4}\frac{\partial}{\partial r}\left[r^4  \left( \frac{1}{6}S_{r\;r\;r}^{\mathbf{u}\mathbf{u} \mathbf{u}}(r,t)- 
    C_{r\;r\;r}^{\mathbf{h}\mathbf{h} \mathbf{u}}(r,t) \right. \right.\\ 
    &
    \left. \left.+  \frac{\partial}{\partial r} \left[\nu S_{r\;r}^{\mathbf{u}\mathbf{u}}(r,t)+
     \lambda  S_{r\;r}^{\mathbf{h}\mathbf{h}}(r,t)\right]\right)\right] +\frac{4}{r} A_{[rt]t}^{\mathbf{u}\mathbf{h} \mathbf{h}}(r,t)\;.
      \label{eq:balance}
\end{align}
\end{widetext}
In the following, we assume statistical stationarity, which allows us to set the l.h.s of Eq. (\ref{eq:balance}) to zero. Multiplying by $r^4$ and integrating from $0$ to $r$ yields
\begin{align}\nonumber
    \lefteqn{S_{r\;r\;r}^{\mathbf{u}\mathbf{u} \mathbf{u}}(r)-6
    C_{r\;r\;r}^{\mathbf{h}\mathbf{h} \mathbf{u}}(r) -\frac{24}{r^4} \int_0^{r}  \textrm{d}r' r'^3  A_{[rt]\;t}^{\mathbf{u}\mathbf{h} \mathbf{h}}(r') } \\
    &=  -\frac{4}{5}
    \langle \varepsilon_{tot} \rangle r + \frac{\partial}{\partial r}\left[\nu S_{r\;r}^{\mathbf{u}\mathbf{u}}(r)+
     \lambda  S_{r\;r}^{\mathbf{h}\mathbf{h}}(r)\right]\;.
     \label{eq:4/5}
\end{align}
In the inertial range, i.e., for $r$ smaller than integral- and larger than dissipation-length scales, we neglect the viscous terms in square brackets and obtain a relation between third-order statistics on the l.h.s. and a term that is proportional to $r$ on the r.h.s. whose magnitude is determined by the total averaged local energy dissipation rate $\langle \varepsilon_{tot} \rangle$~\cite{Friedrich_2016}. 
Nonetheless, in contrast to the hydrodynamic case (\ref{eq:4/5_hydro}) and to the original derivation by Politano and Pouquet~\cite{Politano1998,Politano1998b} where the energy transfer is purely local in scale $r$, we obtain a scaling law in the MHD inertial range (\ref{eq:4/5}) that exhibits a \emph{nonlocal term} stemming from the magnetic induction equation and is a direct consequence of the anti-symmetric tensorial form (\ref{eq:antisym}). As shown in Appendix~\ref{app:recast}, we can re-cast this nonlocal term as
\begin{table*}[ht!]
\centering
\begin{tabular}{c c c c c c c c c c c c c c c }
     \hline
     $\textrm{Re}_{\lambda, kin}$ & $\textrm{Re}_{\lambda, mag}$ & $u_{rms}$ & $h_{rms}$ & $\langle \varepsilon_{kin} \rangle$ & $\langle \varepsilon_{mag} \rangle$ & $\nu= \lambda$ & $\eta_{kin}$ & $\eta_{mag}$ & $\lambda_{kin}$ & $\lambda_{mag}$ & $L_{kin}$ & $L_{mag}$ & $\textrm{d}x$ & N \\
     \hline
     252 & 161 & 2.43 & 1.40 & 2.30 &  5.65& $1.2 \cdot 10^{-3}$ &  $5.2 \cdot 10^{-3}$ & $4.2 \cdot 10^{-3}$ & 0.216 & 0.079 & 1.350 & 0.607 & $6.1\cdot 10^{-3}$ & $1024^3$\\ 
     \hline
     418 & - & 0.686 & - & 0.103 &- & $1.85 \cdot 10^{-4}$ & $2.8 \cdot 10^{-4}$ & - & 0.113 & - & 1.364 & - &  $6.1\cdot 10^{-3}$ & $1024^3$\\
     \hline
\end{tabular}
\caption{Characteristic parameters of the direct numerical simulations (DNS) of 3D MHD and hydrodynamic turbulence. Taylor-based Reynolds numbers $\textrm{Re}_{\lambda, kin}= u_{rms} \lambda_{kin}/\nu$ and $\textrm{Re}_{\lambda, mag}=h_{rms} \lambda_{mag}/\lambda$, root mean square velocity $u_{rms}=\sqrt{\langle {\bf u}^2 \rangle/3}$ and magnetic field $h_{rms}=\sqrt{\langle {\bf h}^2 \rangle/3}$, averaged kinetic and magnetic energy dissipation rates (\ref{eq:eps_kin} and (\ref{eq:eps_mag}), kinematic viscosity $\nu$ and magnetic diffusivity $\lambda$, Kolmogorov dissipation length scales $\eta_{kin}=(\nu^3/ \langle \varepsilon_{kin} \rangle)^{1/4}$ and  $\eta_{mag}=(\lambda^3/ \langle \varepsilon_{mag} \rangle)^{1/4}$, Taylor length scales $\lambda_{kin}= \sqrt{15\nu u_{rms}^2/\langle \varepsilon_{kin} \rangle}$ and $\lambda_{mag}=\sqrt{15\lambda h_{rms}^2/\langle \varepsilon_{mag} \rangle}$, and integral length scales $L_{kin}$ and $L_{mag}$ (both determined from the correlation functions), grid spacing $\textrm{d}x$, and number of grid points $N$. Statistical quantities in the MHD and the hydrodynamic turbulence simulation were averaged over approximately 9 and 5 large eddy turnover times, respectively.}
\label{tab:charac}
\end{table*}
\begin{equation}
  \int_0^{r}  \textrm{d}r' r'^3  A_{[rt]\;t}^{\mathbf{u}\mathbf{h} \mathbf{h}}(r') =- \int_r^{\infty}   \textrm{d}r' r'^3  A_{[rt]\;t}^{\mathbf{u}\mathbf{h} \mathbf{h}}(r') \;,
\end{equation}
which thus represents the influence of magnetic field structures induced at larger scales $r'$ on the local fluctuations at scale $r$ in the sense of the Alfv\'en effect~\cite{alfven,biskamp,iroshnikov1964turbulence,kraichnan1965inertial}. We will now consider further implications of this additional nonlocality for the conservation of energy at small scales and the phenomenon of solar wind heating by turbulent cascades.
In more detail, for small scales $r$ in the vicinity of the dissipation range, nonlinear transfer terms on the l.h.s. in Eq. (\ref{eq:4/5}) should decay faster than the terms on the r.h.s.~\cite{monin}. From a Taylor expansion around $r=0$, we obtain $S_{r\;r\,r}^{\mathbf{u}\mathbf{u} \mathbf{u}}(r) \sim r^3$ as well as
\begin{align}
\label{eq:c_hhu}
C_{r\;r\;r}^{\mathbf{h}\mathbf{h} \mathbf{u}}(r)=& \left. \frac{\partial C_{r\;r\;r}^{\mathbf{h}\mathbf{h} \mathbf{u}}(r)}{\partial r}  \right|_{r=0} r+ \textrm{h.o.t.}\\
A_{[rt]t}^{\mathbf{u}\mathbf{h} \mathbf{h}}(r) =& \left. \frac{\partial A_{[rt]r}^{\mathbf{u}\mathbf{h} \mathbf{h}}(r)}{\partial r}  \right|_{r=0}r+ \textrm{h.o.t.}
\label{eq:a_uhh}
\end{align}
\begin{figure}[ht!]
    \centering
    \includegraphics[width=0.49 \textwidth]{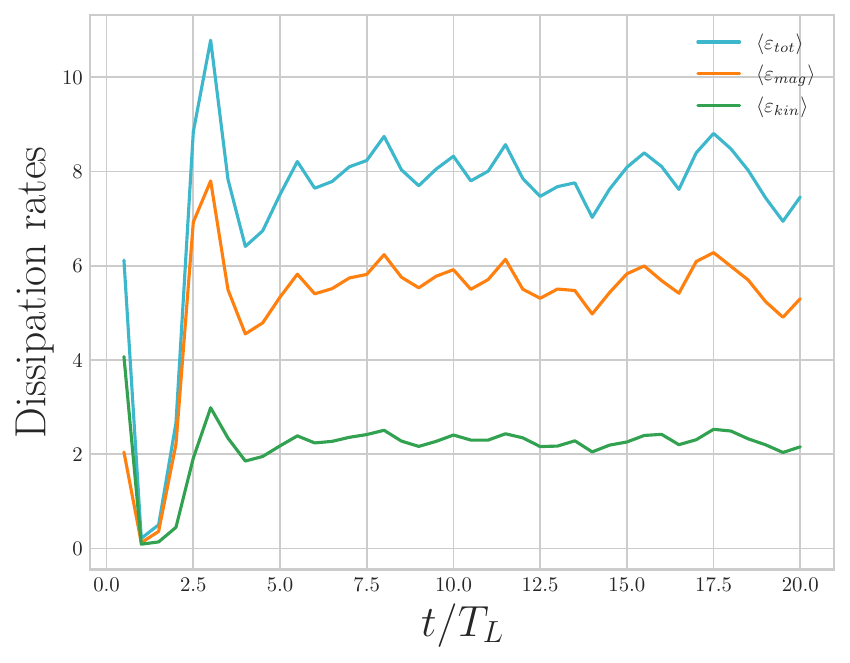}
    \caption{Temporal evolution of the kinetic, magnetic, and total energy dissipation rates throughout the simulation. After an initial oscillation, the dissipation rates remain fairly constant due to the applied forcing mechanism, which indicates a close-to-stationary MHD flow. The assessed statistical quantities in Tab.~\ref{tab:charac} were averaged for $t/T_L>5$. }
    \label{fig:temp_diss}
\end{figure}
\begin{figure}[ht!]
    \centering
    \includegraphics[width=0.495 \textwidth]{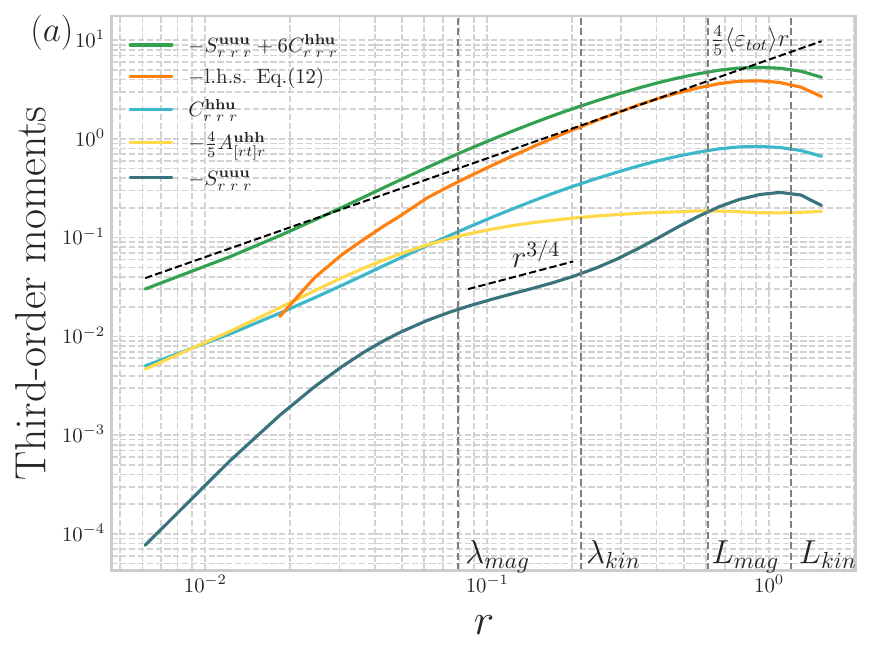}
    \includegraphics[width=0.498 \textwidth]{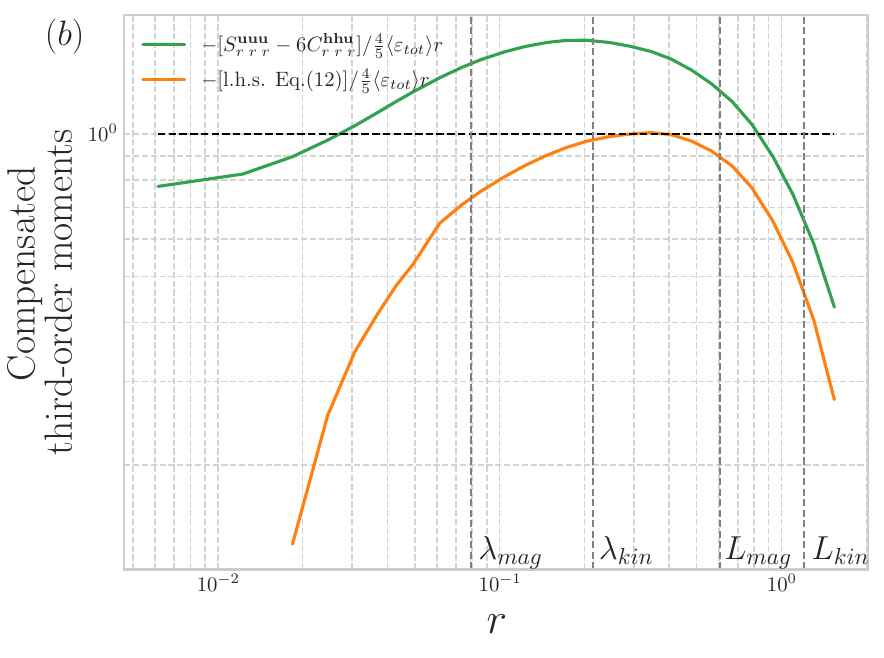}
    \caption{(a) Assessment of the four-fifths law in MHD turbulence from DNS ($1024^3$) of forced MHD turbulence simulation. The dashed line represents the r.h.s. of Eq. (\ref{eq:4/5}, which is determined from the total energy dissipation rate $\varepsilon_{tot}$ of the snapshot. The top curve corresponds the original P\&P law $S_{r\;r\;r}^{\mathbf{u}\mathbf{u} \mathbf{u}}(r)-6
    C_{r\;r\;r}^{\mathbf{h}\mathbf{h} \mathbf{u}}(r)$  derived in~\cite{Politano1998,Politano1998} whereas the additional nonlocal term in Eq. (\ref{eq:4/5}) is closer to the dashed line. The mixed tensors verify the dissipation range behavior (\ref{eq:rel}). The lowest curve represents the third-order longitudinal velocity structure function $S_{r\;r\;r}^{\mathbf{u}\mathbf{u} \mathbf{u}}(r)$. (b) Same as in (a) but compensated by the r.h.s. in Eq. (\ref{eq:4/5}). The original P\&P law (top curve) overestimates the total energy dissipation rate $\varepsilon_{tot}$ by a factor of 1.71 and peaks in front of the inertial range, whereas the inclusion of the nonlocal term in Eq. (\ref{eq:4/5}) leads to the correct prediction (dashed line). }
    \label{fig:4_5}
\end{figure}
Hence, for small $r$, the third-order velocity contribution can be neglected. In contrast, the two other correlation functions (\ref{eq:c_hhu}) and (\ref{eq:a_uhh}) would yield a finite contribution to the nonlinear transfer in the dissipation range (please note that the l.h.s in Eq. (\ref{eq:4/5}) scales as $r$ as well). As was first derived by Chandrasekhar, for small $r$, both correlation tensors are related by
\begin{equation}
    \left. \frac{\partial A_{[rt]r}^{\mathbf{u}\mathbf{h} \mathbf{h}}(r)}{\partial r}  \right|_{r=0}=-\frac{5}{4} \left.
    \frac{\partial C_{r\;r\;r}^{\mathbf{h}\mathbf{h} \mathbf{u}}(r)}{\partial r}  \right|_{r=0}\;,
    \label{eq:rel}
\end{equation}
which is derived in Appendix~\ref{app:relation}. Inserting this relation into Eq. (\ref{eq:4/5}) leads to the exact cancelation of both third-order correlations and reproduces the expected result in the dissipation range, i.e.,
\begin{equation}
    0=-\frac{4}{5}
    \langle \varepsilon_{tot} \rangle r + \frac{\partial}{\partial r}\left[\nu S_{r\;r}^{\mathbf{u}\mathbf{u}}(r)+
     \lambda  S_{r\;r}^{\mathbf{h}\mathbf{h}}(r)\right]\;,
\end{equation}
or
\begin{equation}
    S_{r\;r}^{\mathbf{h}\mathbf{h}}(r)+ \textrm{Pm} S_{r\;r}^{\mathbf{u}\mathbf{u}}(r)= \frac{ \langle \varepsilon_{tot} \rangle}{15 \lambda}
    r^2\;,
\end{equation}
where $\textrm{Pm}={\nu}/{\lambda}$ denotes the magnetic Prandtl number. In the following section, we will verify the modified scaling law (\ref{eq:4/5}) in direct numerical simulations of MHD turbulence.

\section{Verification of the third-order law via direct numerical simulations of MHD turbulence}
\label{sec:dns}
To verify Eq. (\ref{eq:4/5}), we performed DNS of homogeneous and isotropic MHD turbulence in a box of $1024^3$ grid points with characteristic turbulence parameters summarized in Tab.~\ref{tab:charac}. The forcing scheme consists of a divergence-free random forcing~\cite{Alvelius_1999} that is applied to the evolution equation of the velocity field (i.e., the evolution equation for the magnetic field is not actively forced). As shown in Fig.~\ref{fig:temp_diss}, after an initial phase, the forcing scheme leads to nearly constant kinetic and magnetic energy dissipation rates $\langle \varepsilon_{kin} \rangle$ and $\langle \varepsilon_{mag} \rangle$ over time. Fig. \ref{fig:4_5} depicts the third-order moments that enter Eq. (\ref{eq:4/5}). The lowest curve corresponds to the third-order velocity structure function $S_{r\;r\;r}^{\mathbf{u}\mathbf{u} \mathbf{u}}(r)$. Interestingly, this curve exhibits three distinct scaling regions: In the dissipation range, we observe the predicted scaling $\sim r^3$ followed by a region that scales in agreement with the Iroshnikov-Kraichnan phenomenology $\sim r^{3/4}$ whereas a linear increase seems to dominate at larger $r$. Furthermore, in the dissipation range, we can directly verify the relation between the symmetric and anti-symmetric correlations (\ref{eq:rel}). In the case under study, the original prediction~\cite{Politano1998,Politano1998b} without the additional nonlocal source term overestimates the total energy dissipation rate $\langle \varepsilon_{tot} \rangle$ (dashed line) and does not vanish for small $r$, therefore, violating energy conservation. This becomes even more apparent in the compensated plot in Fig.~\ref{fig:4_5} (b), which suggests that including the source term results in the correct scaling in the inertial range in between the largest Taylor scale $\lambda_{kin}$ and the smallest integral length scale $L_{mag}$. In contrast, the scaling obtained using the P\&P law~\cite{Politano1998,Politano1998b} (green curve) would plateau before this inertial range and overestimate the $\langle \varepsilon_{tot} \rangle$ by a factor of 1.71. 
We must emphasize that this discrepancy between the original P\&P law and (\ref{eq:4/5}) might have been barely detectable in previous numerical studies of MHD turbulence~\cite{mininni2009finite,sorriso2002analysis,yousef2007exact} due for instance to low magnetic and fluid Reynolds numbers and signatures of the large-scale anisotropic forcing scheme (e.g., by Taylor-Green vortices). Indeed, MHD and kinetic simulations reproduce plasma dynamics and features peculiar to the set of governing parameters defined for specific runs, which may differ significantly from actual solar wind parameters. Due to the insufficient computational power of today's supercomputers, constraints in reproducing the solar wind numerically are technical but also inherent since solar wind dynamics develop in a vast parameter space, hardly reproducible even with large ensembles of simulations.

This entails that the influence of the additional nonlocal term on cascade processes can be sub-dominant to the third-order correlation stemming from the Lorentz force $C_{r\;r\;r}^{{\bf h}{\bf h}{\bf u}}({\bf r},t)$ or that the assumptions of isotropy and homogeneity in the derivation of Eq. (\ref{eq:4/5}) are not strictly guaranteed, in some of the simulations appearing in the literature, and also in many cases within the actual solar wind. Nonetheless, for some settings, e.g., strong Alfv\'enic turbulence, this term might play an important, non-negligible role.
\section{Conclusions}
\label{sec:conc}
In this paper, we have derived from first principles an additional nonlocal term in the equation for the energy cascade directly from the incompressible MHD equations, assuming isotropy and homogeneity only. This additional term can potentially improve the heating rate predictions by the original P\&P law when nonlocal interactions become essential in the overall plasma dynamics. We have verified by numerical simulations of  MHD turbulence forced at large scales that this additional nonlocal term is critical in describing the energy transfer from large to small scales for the set of parameters chosen.

The exact law established by Politano and Pouquet~\cite{Politano1998,Politano1998b} assumes homogeneity, isotropy, and incompressibility - conditions that are fulfilled perhaps only locally in interplanetary space plasmas - representing to date the most robust statistical framework to characterize energy transfer and heating in the solar wind~\cite{MARINO2023,Marino_2008}.
Without diminishing its validity, our findings suggest that in some solar wind samples, the current analyses of observational data by the original P\&P law and some of its surrogates might substantially overestimate the rate at which energy is dissipated (or further transferred across sub-proton scales) at the bottom of the MHD turbulent cascade developing in the solar wind. This leads to improved third-order moment law predictions characterizing the observed space plasma heating. 

As mentioned in Sec.~\ref{sec:derivation}, the additional nonlocal term arises due to the advection of the magnetic field by the velocity field and the corresponding transformation of the electric field ${\bf E}' ={\bf E}+ \mu {\bf u} \times {\bf H}$. Hence, it would also be interesting to assess other transformations of the electric field, e.g., in the context of Hall MHD~\cite{galtier2008karman,Foldes2023}.
Future work will be devoted to applying the law (\ref{eq:4/5}) and assessing the influence of nonlocality in solar wind measurements. Moreover, we aim at reconstructing velocity and magnetic fields from partial measurements by extending current velocity field reconstructions~\cite{Friedrich:2022aa,lubke2023stochastic} to MHD turbulence~\cite{friedrich2020probability,friedrich2020generalized,lubke2024towards}, exploiting observations of state-of-art ongoing \citep{SolO,PSP} and future multi-spacecraft solar wind mission \citep{Helioswarm,Pecora1,Pecora2,Plasma_obs}. 


\section{Acknowledgments}
J.F. acknowledges fruitful discussions with H. Homann, A. Pumir, and J. Peinke.
We gratefully acknowledge the Gauss Centre for Supercomputing 
(www.gauss-centre.eu) for funding this project by providing computing time through the John von Neumann Institute for Computing (NIC) on the GCS Supercomputer JUWELS \cite{jsc-juwels:2021} at Jülich Supercomputing Centre (JSC). The simulation output was also analyzed on HPC facilities at the \'Ecole Centrale de Lyon (PMCS2I) in Ecully (France) and the HPC Cluster STORM, located at the University of Oldenburg (Germany) and funded through the REACT-EU program by the NBank and the Ministry of Science and Culture (MWK) of the Lower Saxony State under grant number ZW7-85186744.
%
R.M. acknowledges support from the project ``EVENTFUL'' (ANR-20-CE30-0011), funded by the French ``Agence Nationale de la Recherche'' - ANR through the program AAPG-2020. M.W.  acknowledges support from the German Science Foundation DFG within the Collaborative Research Center SFB1491.

\appendix
\section{Derivation of the Equation of Total Energy Balance in MHD Turbulence}
\label{app:energy_balance}
In this section, we derive an evolution equation for the total energy (i.e., the sum of kinetic and magnetic energy) in MHD turbulence. Therefore, we consider the MHD equations
\begin{widetext}
\begin{align}\label{eq:u}
  \frac{\partial}{\partial t}u_i({\bf x},t) + u_k({\bf x},t) \frac{\partial}{\partial x_k} u_i({\bf x},t) -h_k({\bf x},t)
\frac{\partial}{\partial x_k}h_i({\bf x},t) &= -\frac{1}{\rho} \frac{\partial}{\partial x_i}P({\bf x},t) + \nu \nabla^2_{\bf x} u_i({\bf x},t) \;,
\\
\frac{\partial }{\partial t}h_i({\bf x},t) + u_k({\bf x},t) \frac{\partial}{\partial x_k} h_i({\bf x},t) -h_k({\bf x},t)
\frac{\partial}{\partial x_k}u_i({\bf x},t) &= \lambda \nabla^2_{\bf x} h_i({\bf x},t) \;,
\label{eq:h}
\end{align}
\end{widetext}
where summation over identical indices is implied. Here, $P({\bf x},t)=p({\bf x},t)+\frac{\rho}{2} |{\bf h}|^2$ denotes the sum of hydrodynamic pressure $p({\bf x},t)$ and 
the magnetic pressure, $\rho$ the density (which will be set to one), $\nu$ the kinematic
viscosity and $\lambda$ the magnetic diffusivity of the conducting fluid. Furthermore, it should be noted that in this formulation of the MHD equations, the magnetic field ${\bf H}$ is measured in units of the velocity field
\begin{equation}
 {\bf h}({\bf x},t) = \sqrt{\frac{\mu}{4\pi \rho}}  {\bf H}({\bf x},t)\;.
\end{equation}

The total energy can now be defined as the sum of kinetic and magnetic energy according to
\begin{equation}
    E_{tot}(t) = \frac{1}{2} \left \langle u_i u_i + h_i h_i\right \rangle 
    =\frac{3}{2} \left(u_{rms}^2 + h_{rms}^2 \right)\;,
\end{equation}
where $u_{rms}$ and $h_{rms}$ denote the root mean square velocity and magnetic field of a three-dimensional MHD flow. Moreover, the brackets $\langle \ldots \rangle$ represent suitable averages, e.g., ensemble averages. We scalar multiply Eq. (\ref{eq:u}) by $u_i$ and Eq. (\ref{eq:h}) by $h_i$
\begin{align}\label{eq:uu_app}\
  \frac{1}{2}\frac{\partial}{\partial t}{\bf u}^2 + \frac{1}{2} \frac{\partial}{\partial x_k} u_k {\bf u}^2 - u_i\frac{\partial}{\partial x_k} h_k h_i =& -\frac{1}{\rho} \frac{\partial}{\partial x_i} u_i P\\ \nonumber
  &+ u_i \nu \nabla^2_{\bf x} u_i \;,
\\
 \frac{1}{2}\frac{\partial}{\partial t}{\bf h}^2+ \frac{1}{2} \frac{\partial}{\partial x_k} u_k {\bf h}^2 - h_i\frac{\partial}{\partial x_k} h_k u_i  =&  h_i \lambda \nabla^2_{\bf x} h_i  \;,
\label{eq:hh_app}
\end{align}
where we used the incompressibility conditions for both velocity and magnetic field, i.e., $\frac{\partial}{\partial x_k} u_k=0$ and $\frac{\partial}{\partial x_k}h_k=0$. Here, the viscous terms can be reformulated according to 
\begin{align} \nonumber
 \lefteqn{u_i \nu \nabla^2_{\bf x} u_i = \nu \frac{\partial}{\partial x_k} u_i \frac{\partial u_i}{\partial x_k} -  \nu\sum_{i,k} \left(\frac{\partial u_i}{\partial x_k} \right)^2}\\ \nonumber
 =& \frac{\nu}{2} \nabla^2_{\bf x} {\bf u}^2  -  \frac{\nu}{2}\sum_{i,k} \left(\frac{\partial u_i}{\partial x_k} +\frac{\partial u_k}{\partial x_i} \right)^2+\nu \sum_{i,k} \left(\frac{\partial u_i}{\partial x_k}\frac{\partial u_k}{\partial x_i} \right)\\
 =& \frac{\nu}{2} \nabla^2_{\bf x} {\bf u}^2 - \varepsilon_{kin} + \nu \frac{\partial}{\partial x_i} u_k \frac{\partial}{\partial x_k} u_i\;.
\end{align}
The same procedure applies to the diffusive terms in Eq. (\ref{eq:hh_app}). Furthermore, local kinetic and magnetic energy dissipation rates are defined according to
\begin{align}
     \varepsilon_{kin}({\bf x},t) =& \frac{\nu}{2} \sum_{i,k} \left(\frac{\partial u_i({\bf x},t)}{\partial x_k} +\frac{\partial u_k({\bf x},t)}{\partial x_i} \right)^2\;, \\
     \varepsilon_{mag}({\bf x},t) =& \frac{\lambda}{2} \sum_{i,k} \left(\frac{\partial h_i({\bf x},t)}{\partial x_k} +\frac{\partial h_k({\bf x},t)}{\partial x_i} \right)^2\;.
\end{align}
We observe that nonlinear terms which are advected by the magnetic field  (third terms on l.h.s of Eqs. (\ref{eq:uu_app}) and (\ref{eq:hh_app})) can be cast in conservative form by adding Eqs. (\ref{eq:uu_app}) and (\ref{eq:hh_app}), which yields
\begin{equation}
    \frac{\partial}{\partial t} e_{tot}({\bf x},t)+ \nabla_{{\bf x}} \cdot {\bf J}^{tot}({\bf x},t)= - \varepsilon_{tot}({\bf x},t)\;,
\end{equation}
where we defined
\begin{align}
    &e_{tot}({\bf x},t) = \frac{{\bf u}^2({\bf x},t)+{\bf h}^2({\bf x},t)}{2}\;,\\ \nonumber
    \label{eq:j_tot}
    &{\bf J}^{tot}({\bf x},t) = {\bf u}({\bf x},t)\left(\frac{{\bf u}^2({\bf x},t)+{\bf h}^2({\bf x},t)}{2}+ \frac{P({\bf x},t) }{\rho}\right) \\ \nonumber
    &- {\bf h}({\bf x},t) [{\bf u}({\bf x},t)\cdot {\bf h}({\bf x},t)] 
     -\frac{\nu}{2} \nabla_{\bf x} {\bf u}^2 ({\bf x},t) -  \frac{\lambda}{2} \nabla_{\bf x} {\bf h}^2 ({\bf x},t)\\ 
     &- \nu {\bf u}({\bf x},t) \cdot \nabla_{{\bf x}}  {\bf u}({\bf x},t)  -\lambda  {\bf h}({\bf x},t) \cdot \nabla_{{\bf x}}  {\bf h}({\bf x},t) \;,\\
    &\varepsilon_{tot}({\bf x},t) =  \varepsilon_{kin}({\bf x},t) +  \varepsilon_{mag}({\bf x},t)\;. 
\end{align} 
The two first terms in the brackets of (\ref{eq:j_tot} denote the kinetic and magnetic energy transported throughout the surface. The second term is the work from the fluid against the total pressure. The next term is the cross helicity ${\bf u}({\bf x},t)\cdot {\bf h}({\bf x},t)$ which is transported anti-parallel along the lines of force. Performing the averaging procedure, we obtain a temporal evolution equation for the total energy $E_{tot}(t)= \langle e_{tot}({\bf x},t) \rangle$ as
\begin{equation}
\dot E_{tot}(t)=  -\langle \varepsilon_{tot}({\bf x},t)\rangle \;,
\label{eq:e_tot}
\end{equation}
where we assumed homogeneity to neglect the transport terms $\nabla_{\bf x} \cdot \langle  {\bf J}^{tot}({\bf x},t) \rangle =0$. Hence, the total energy is only changed by viscous and magnetic dissipation. Here, we only considered decaying MHD turbulence; otherwise, the temporal evolution $E_{tot}(t)$ would be influenced by additional forcing mechanisms in Eqs. (\ref{eq:u}) and (\ref{eq:h}) that would lead to additional source terms on the r.h.s. of Eq. (\ref{eq:e_tot}).
\section{Derivation of the Friedmann-Keller Hierarchy in Homogeneous and Isotropic MHD Turbulence}
\label{app:friedmann}
In this section, we recapitulate Chandrasekhar's derivation~\cite{chandra:1951} of the Friedmann-Keller hierarchy~\cite{keller1924}
for magnetic and velocity field correlation functions. In a statistical description of MHD turbulence,
evolution equations for the two-point correlation tensors
\begin{align}\label{eq:c_uu}
     C_{i\;j}^{{\bf u}{\bf u}}({\bf r}, t)=& \left \langle u_i ({\bf x},t) u_j({\bf x}',t) \right \rangle\;,
     \\
     C_{i\;j}^{{\bf u}{\bf h}}({\bf r}, t)=& \left \langle u_i ({\bf x},t) h_j({\bf x}',t) \right \rangle\;, \label{eq:c_uh}\\
     C_{i\;j}^{{\bf h}{\bf h}}({\bf r}, t)=& \left \langle h_i ({\bf x},t) h_j({\bf x}',t) \right \rangle\;,
     \label{eq:c_hh}
\end{align}
are obtained by the same procedure as in the hydrodynamic case~\cite{DeKarman1938} and were first derived by Chandrasekhar~\cite{chandra:1951}. In the following, we are solely interested in evolution equations for the two-point velocity and magnetic field tensors (\ref{eq:c_uu}) and (\ref{eq:c_hh}). The evolution equation and invariant form of the cross helicity tensor (\ref{eq:c_uh}), which is a skew-symmetric tensor, can be found in~\cite{chandra:1951}. The procedure for the velocity correlation tensor (\ref{eq:c_uu}) starts with multiplying Eq. (\ref{eq:u}) by $u_j'=u_j({\bf x}',t)$
\begin{align}\nonumber
     \lefteqn{u_j'\frac{\partial}{\partial t}u_i + \frac{\partial}{\partial x_k} \left(u_k u_i u_j' - 
    h_k h_i u_j'\right)} \\ 
    &= -u_j'\frac{1}{\rho} \frac{\partial}{\partial x_i} P+ \nu u_j'  \nabla^2_{\bf x} u_i\;,
    \label{eq:u_i}
\end{align}
where we used the incompressibility condition for both velocity and magnetic field, i.e., $\frac{\partial}{\partial x_k} u_k=0$ and $\frac{\partial}{\partial x_k} h_k=0$. In the same manner, we can multiply the evolution equation for $u_j({\bf x}',t)$ by $u_i=u_i({\bf x},t)$, which yields
\begin{align}\nonumber
     \lefteqn{u_i\frac{\partial}{\partial t}u_j' + \frac{\partial}{\partial x_k'} \left(u_k' u_j' u_i - 
    h_k' h_j' u_i\right) }\\ 
    &= -u_i\frac{1}{\rho} \frac{\partial}{\partial x_i'} P'+ \nu u_i \nabla^2_{{\bf x}'} u_j'\;.
    \label{eq:u_j}
 \end{align}
Moreover, from the induction equation (\ref{eq:h}), we obtain
\begin{equation}
 h_j'\frac{\partial}{\partial t}h_i + \frac{\partial}{\partial x_k} \left(u_k h_i h_j' -
    h_k u_i h_j'\right) = \lambda h_j'  \nabla^2_{\bf x} h_i\;,
    \label{eq:h_i}
\end{equation}
as well as
\begin{equation}
     h_i\frac{\partial}{\partial t}h_j' + \frac{\partial}{\partial x_k'} \left(u_k' h_j' h_i - 
    h_k' u_j' h_i \right) = \lambda h_i \nabla^2_{{\bf x}'} h_j'\;.
    \label{eq:h_j}
\end{equation}
Before we add Eqs. (\ref{eq:u_i}) and (\ref{eq:u_j}) and take the ensemble average $\langle \ldots \rangle$, we discuss certain simplifications that are a direct consequence of the assumptions of homogeneity and isotropy of the MHD flow:
\begin{itemize}
 \item Based on the assumption of homogeneity, correlation functions solely depend on the relative distance ${\bf r}= {\bf x}' -{\bf x}$, and we obtain
\begin{align} \label{eq:corr0}
 & \quad C_{i\;j}^{{\bf u}{\bf u}}({\bf r}, t)= \langle u_i ({\bf x},t) u_j({\bf x}',t) \rangle =
C_{i\;j}^{{\bf u}{\bf u}}(-{\bf r}, t)\;,
\\
& \quad C_{(ki)j}^{{\bf u}{\bf u}{\bf u}}({\bf r}, t)= \langle u_k({\bf x},t) u_i ({\bf x},t) u_j({\bf
x}',t) \rangle\;, \\
& \quad C_{(kj)i}^{{\bf u}{\bf u}{\bf u}}(-{\bf r}, t)= \langle u_k({\bf x}',t) u_j ({\bf x}',t) u_i({\bf
x},t) \rangle\;.
\label{eq:corr1}
\end{align}
Therefore, viscous terms can be rewritten according to
\begin{align} \nonumber
\lefteqn{ [\nabla^2_{\bf x} + \nabla^2_{{\bf x}'}] \left \langle u_i ({\bf x},t) u_j({\bf x}',t) \right \rangle}\\
=& 2
\nabla^2_{\bf r}
\langle u_i ({\bf x},t) u_j({\bf x}',t) \rangle= 2\nabla^2_{\bf r}C_{i\;j}^{{\bf u}{\bf u}}({\bf r}, t)\;.
\end{align}
Correlations where the magnetic field occurs an even number of times transform identically to Eqs. (\ref{eq:corr0}-\ref{eq:corr1})
\begin{align} \label{eq:hcorr0}
 & \;\;\;\quad C_{i\;j}^{{\bf h}{\bf h}}({\bf r}, t)= \langle h_i ({\bf x},t) h_j({\bf x}',t) \rangle =
C_{i\;j}^{{\bf h}{\bf h}}(-{\bf r}, t)\;,\\
\label{eq:hcorr1}
& \;\;\;\quad C_{(ki)j}^{{\bf h}{\bf h}{\bf u}}({\bf r}, t)= \langle h_k({\bf x},t) h_i ({\bf x},t) u_j({\bf
x}',t) \rangle\;, \\
& \;\;\;\quad C_{(kj)i}^{{\bf h}{\bf h}{\bf u}}(-{\bf r}, t)=\langle h_k({\bf x}',t) h_j ({\bf x}',t) u_i({\bf
x},t) \rangle\;,
\end{align}
and for the anti-symmetric tensor, we obtain
\begin{align}
\label{eq:ant}
&A_{[ki]j}^{{\bf u}{\bf h}{\bf h}}({\bf r}, t) \\ \nonumber
\quad =& \langle (u_k({\bf x},t) h_i ({\bf x},t)-u_i({\bf x},t) h_k ({\bf x},t) )h_j({\bf
x}',t)  \rangle\;, \\
& A_{[kj]i}^{{\bf u}{\bf h}{\bf h}}(-{\bf r}, t) \\ \nonumber
 \quad =& \langle (u_k({\bf x}',t) h_j ({\bf x}',t)-u_j({\bf x}',t) h_k ({\bf x}',t) )h_i({\bf
x},t) \rangle\;.
\end{align}
Here, square brackets in the index of Eq. (\ref{eq:ant}) indicate that the corresponding tensor is anti-symmetric in $i$ and $k$.
\item Furthermore, isotropic and mirror-symmetric tensors of the third order obey the following relation
\begin{align}
 C_{(kj)i}^{{\bf u}{\bf u}{\bf u}}(-{\bf r}, t) =& -C_{(kj)i}^{{\bf u}{\bf u}{\bf u}}({\bf r}, t)\;, \\ 
 C_{(kj)i}^{{\bf h}{\bf h}{\bf u}}(-{\bf r}, t) =& -C_{(kj)i}^{{\bf h}{\bf h}{\bf u}}({\bf r}, t)\;, 
 \label{eq:iso}
 \end{align}
 and for the anti-symmetric tensor
 \begin{align}
 A_{[kj]i}^{{\bf u}{\bf h}{\bf h}}(-{\bf r}, t)=-A_{[kj]i}^{{\bf u}{\bf h}{\bf h}}({\bf r}, t)\;.
\end{align}
\item Pressure-velocity contributions vanish based on isotropy and mirror symmetry~\cite{Faust2015}, which yields
\begin{equation}
\left(\frac{\partial}{\partial x_i} + \frac{\partial}{\partial x_i'} \right )
\langle P({\bf x},t) u_j({\bf x}',t) \rangle=0\;.
\end{equation}
\end{itemize}
Under the above assumptions, we obtain from Eqs. (\ref{eq:u_i}) and (\ref{eq:u_j})
\begin{align} \nonumber
& \frac{\partial}{\partial t} C_{i\;j}^{{\bf u}{\bf u}}({\bf r}, t)- \frac{\partial}{\partial r_k} \left(
C_{(ki)j}^{{\bf u}{\bf u}{\bf u}}({\bf r}, t)+C_{(kj)i}^{{\bf u}{\bf u}{\bf u}}({\bf r}, t) \right. \\
&\left. -C_{(ki)j}^{{\bf h}{\bf h}{\bf u}}({\bf r}, t)-C_{(kj)i}^{{\bf h}{\bf h}{\bf u}}({\bf r}, t)\right) = 2\nu \nabla^2_{\bf r}  C_{i\;j}^{{\bf u}{\bf u}}({\bf r}, t)\;.
\end{align}
Furthermore, as terms like $\frac{\partial}{\partial r_k} C_{(ki)j}^{{\bf u}{\bf u}{\bf u}}({\bf r}, t)$ are second order isotropic tensors, they must be symmetric in $i$ and $j$. We thus obtain
\begin{align}\nonumber
 &\frac{\partial}{\partial t} C_{i\;j}^{{\bf u}{\bf u}}({\bf r}, t)- 2\frac{\partial}{\partial r_k} \left(
C_{(ki)j}^{{\bf u}{\bf u}{\bf u}}({\bf r}, t)-C_{(ki)j}^{{\bf h}{\bf h}{\bf u}}({\bf r}, t)\right) \\
=& 2\nu \nabla^2_{\bf r}  C_{i\;j}^{{\bf u}{\bf u}}({\bf r}, t)\;.
\label{eq:u_kar}
\end{align}
The same procedure can be applied to the induction equation (\ref{eq:h}) and we obtain
\begin{equation}
    \frac{\partial}{\partial t} C_{i\;j}^{{\bf h}{\bf h}}({\bf r}, t)- 2 \frac{\partial}{\partial r_k} 
A_{[ki]j}^{{\bf u}{\bf h}{\bf h}}({\bf r}, t) = 2\lambda \nabla^2_{\bf r}  C_{i\;j}^{{\bf h}{\bf h}}({\bf r}, t)\;.
\label{eq:h_kar}
\end{equation}
These two equations (and the additional equation for the cross helicity tensor of second order derived by Chandrasekhar~\cite{chandra:1951}) are a generalization of the Friedmann-Keller hierarchy~\cite{keller1924} to homogeneous and isotropic MHD turbulence.
\section{Derivation of Evolution Equations for Longitudinal Velocity and Magnetic Field Correlation Functions of Second Order}
\label{app:kh}
A further simplification of Eqs. (\ref{eq:u_kar}) and (\ref{eq:h_kar}) can be obtained from the invariant theory of homogeneous and isotropic turbulence~\cite{Robertson1940}. Each tensor can be rewritten in terms of its longitudinal correlation function, which is a consequence of the incompressibility condition and is further derived in Appendices \ref{app:chan},\ref{app:chan_new}, and \ref{app:struc}. The second-order tensors (\ref{eq:corr0}) and (\ref{eq:hcorr0}) both follow the tensorial form
\begin{align} \nonumber
  C_{i j}({\bf r},t) =& \left(C_{rr}(r,t)- \frac{1}{2 r}
\frac{\partial}{\partial r}(r^2 C_{rr}(r,t)) \right )\frac{r_i r_j}{r^2}\\
&+\frac{1}{2r} \frac{\partial}{\partial r} (r^2 C_{rr}(r,t)) \delta_{ij}\;,
\label{eq:ukar}
\end{align}
where $C_{rr}(r,t)$ denotes the longitudinal correlation function of second order. In the same manner, third-order tensors that are symmetric in $i$ and $k$, such as (\ref{eq:corr1}) and (\ref{eq:hcorr1}) obey the following form
\begin{align}\nonumber
  &C_{(ki)j}({\bf r},t)=-\frac{r^2}{2} \frac{\partial}{\partial r}
\left( \frac{ C_{rrr} (r,t)}{r} \right) \frac{r_i r_j r_k}{r^3}\\ \nonumber
 &+ \frac{1}{4 r}\frac{\partial}{\partial r} \left ( r^2 C_{rrr}
(r,t) \right) \left( \frac{r_i}{r} \delta_{kj} + \frac{r_k}{r} \delta_{ij} \right)\\
&-
\frac{C_{rrr} (r,t)}{2} \frac{r_j}{r} \delta_{ik}\;.
\label{eq:ukarneu}
\end{align}
It should be noted that this tensorial form is only applicable to third-order terms in Eq. (\ref{eq:u_kar}). The third-order correlation in Eq. (\ref{eq:h_kar}) is anti-symmetric in $k$ and $i$ and follows a different tensorial form namely
\begin{equation}
     A_{[ki]j}^{\mathbf{u}\mathbf{h}{\mathbf{h}}}({\bf r},t)= A_{[rt]t}^{\mathbf{u}\mathbf{h}{\mathbf{h}}}(r,t) \left(\frac{r_i}{r}\delta_{jk}- \frac{r_k}{r}\delta_{ij}\right)\;.
     \label{eq:anti_sym}
\end{equation}
Summing over $i=j$ in Eqs. (\ref{eq:u_kar}-\ref{eq:h_kar}) yields
\begin{widetext}
\begin{eqnarray}\label{eq:Q_kin}
 Q_{kin}(r,t)&=& \frac{1}{2}\sum_{i=j} C_{i\;j}^{{\bf u}{\bf u}}({\bf r}, t)\;, \quad
J^{kin}_k({\bf r},t) = -\sum_{i=j} \left(C_{(ki)j}^{{\bf u}{\bf u}{\bf u}}({\bf r}, t)-C_{(ki)j}^{{\bf h}{\bf h}{\bf u}}({\bf r}, t)\right)\;,\\
Q_{mag}(r,t)&=& \frac{1}{2}\sum_{i=j} C_{ij}({\bf r},t)\;, \quad
J^{mag}_k({\bf r},t) = -\sum_{i=j} A_{[ki]j}^{{\bf u}{\bf h}{\bf h}}({\bf r}, t)\;,
\end{eqnarray}
\end{widetext}
 and we obtain two balance equation for $Q_{kin}(r,t)$ and $Q_{mag}(r,t)$  with their corresponding currents ${\bf J}^{kin}({\bf r},t)$ and ${\bf J}^{mag}({\bf r},t)$ that read
\begin{align}
 \frac{\partial}{\partial t} Q_{kin}(r,t)  + \nabla_{\bf r} \cdot {\bf J}^{kin}({\bf r},t)=&
2{\nu } \Delta_{\bf r} Q_{kin}(r,t)\;,
\label{eq:ukar_balance}\\
 \frac{\partial}{\partial t} Q_{mag}(r,t) + \nabla_{\bf r} \cdot {\bf J}^{mag}({\bf r},t)=&
2{ \lambda } \Delta_{\bf r} Q_{mag}(r,t)\;.
\label{eq:hkar_balance}
\end{align}
As shown in Appendix~\ref{app:struc}, 
$Q_{kin}(r,t)$, $Q_{mag}(r,t)$ and their corresponding currents ${\bf J}^{kin}({\bf r},t)$, ${\bf J}^{mag}({\bf r},t)$ can be expressed as 
\begin{widetext}
\begin{eqnarray}\label{insert}
 \frac{\partial}{\partial t} Q_{kin}(r,t)&=& \frac{1}{2r^2}\frac{\partial}{\partial r} \left( r^3
\frac{\partial}{\partial t}C_{r\;r}^{{\bf u}{\bf u}}(r,t)  \right), \qquad \frac{\partial}{\partial t} Q_{mag}(r,t)= \frac{1}{2r^2}\frac{\partial}{\partial r} \left( r^3
\frac{\partial}{\partial t}C_{r\;r}^{{\bf h}{\bf h}}(r,t)  \right),\\
 \frac{\partial}{\partial r_k} J_k^{kin}({\bf r},t) &=& -\frac{1}{2r^2} \frac{\partial}{\partial r}
\left( \frac{1}{r} \frac{\partial}{\partial r}\left [ r^4 \left(C_{r\;r\;r}^{{\bf u}{\bf u}{\bf u}}-C_{r\;r\;r}^{{\bf h}{\bf h}{\bf u}} \right) \right] \right)\;, \qquad \frac{\partial}{\partial r_k} J_k^{mag}({\bf r},t) = \frac{2}{r^2}\frac{\partial}{\partial r} \left(r^2 A_{[rt]t}^{\mathbf{u}\mathbf{h}{\mathbf{h}}}(r,t)\right)\;,\\
\Delta_{\bf r} Q_{kin}(r,t) &=& \frac{1}{r^2} \frac{\partial}{\partial r} \left( r^2
\frac{\partial}{\partial r}  Q_{kin}(r,t) \right)\;, \qquad \Delta_{\bf r} Q_{mag}(r,t) = \frac{1}{r^2} \frac{\partial}{\partial r} \left( r^2
\frac{\partial}{\partial r}  Q_{mag}(r,t) \right)\;.
\end{eqnarray}
\end{widetext}
Inserting these relations into Eq. (\ref{eq:ukar_balance}) and  (\ref{eq:hkar_balance}) yields 
\begin{align} \nonumber
    \frac{\partial}{\partial t} C_{r\;r}^{\mathbf{u}\mathbf{u}}(r,t) =& \frac{1}{r^4}\Bigg[\frac{\partial}{\partial r}r^4  \Bigg( C_{r\;r\;r}^{\mathbf{u}\mathbf{u} \mathbf{u}}(r,t)-
    C_{r\;r\;r}^{\mathbf{h}\mathbf{h} \mathbf{u}}(r,t) \\
    &+ 2 \nu \frac{\partial}{\partial r} C_{r\;r}^{\mathbf{u}\mathbf{u}}(r,t) \Bigg) \Bigg]\;, \label{eq:kh_u_app}
     \\
       \frac{\partial}{\partial t} C_{r\;r}^{\mathbf{h}\mathbf{h}}(r,t) =& -\frac{4}{r} A_{[rt]t}^{\mathbf{u}\mathbf{h} \mathbf{h}}(r,t)+ 2 \lambda \frac{1}{r^4} \frac{\partial}{\partial r} r^4 \frac{\partial}{\partial r} C_{r\;r}^{\mathbf{h}\mathbf{h}}(r,t) \;.
       \label{eq:kh_h_app}
\end{align}
Eq. (\ref{eq:kh_u_app}) is the generalization of the von K\'arm\'an-Howarth equation to MHD turbulence with the additional third-order correlation function $C_{r\;r\;r}^{\mathbf{h}\mathbf{h} \mathbf{u}}(r,t) $ that is due to the Lorentz force in Eq. (\ref{eq:u}). The evolution equation for the second-order longitudinal correlation function for the magnetic field (\ref{eq:kh_h_app}) possesses a different differential form, which is due to the anti-symmetry of the tensor in Eq. (\ref{eq:anti_sym}). 
\section{On the Equivalence of the Mixed Third-Order Correlations for Small Scale Separations}
\label{app:relation}
As first recognized by Chandrasekhar~\cite{chandra:1951}, the defining scalars of the mixed third-order correlation functions $C_{(ki)j}^{{\bf h}{\bf h}{\bf u}}({\bf r}, t)= \left \langle h_i({\bf x},t)h_k({\bf x},t)u_j({\bf x}',t) \right \rangle$ and  $A_{[ki]j}^{{\bf u}{\bf h}{\bf h}}({\bf r}, t)= \left \langle (h_i({\bf x},t)u_k({\bf x},t)- h_k({\bf x},t)u_i({\bf x},t))h_j({\bf x}',t) \right \rangle$ are related to each other for small scale separations ${\bf r}$. This can be shown as follows: First, we consider the derivative of the symmetric tensor defined by Eq. (\ref{eq:ukarneu})
\begin{align} \nonumber
    &\left. \frac{\partial C_{(ki)j}^{{\bf h}{\bf h}{\bf u}}({\bf r}, t) }{\partial r_i}\right|_{{\bf r}=0}= \left \langle h_i({\bf x},t) h_k({\bf x},t) \frac{\partial u_j({\bf x},t)}{\partial x_i} \right \rangle \\ \nonumber
    =& \left[ \frac{ C_{r\;r\;r}^{\mathbf{h}\mathbf{h} \mathbf{u}}(r,t)}{r}+ \frac{3}{2} \frac{\partial  C_{r\;r\;r}^{\mathbf{h}\mathbf{h} \mathbf{u}}(r,t)}{\partial r}\right]_{r=0} \delta_{jk} \\ 
    =& \frac{5}{2}\left.\frac{\partial  C_{r\;r\;r}^{\mathbf{h}\mathbf{h} \mathbf{u}}(r,t)}{\partial r}\right|_{r=0} \delta_{jk}\;. 
    \label{eq:ident0}
\end{align}
Similarly, we derive the tensorial form of the anti-symmetric third-order correlation tensor (\ref{eq:anti_sym}) and obtain
\begin{align} \nonumber
    &\left. \frac{\partial A_{[ki]j}^{{\bf u}{\bf h}{\bf h}}({\bf r}, t) }{\partial r_i}\right|_{{\bf r}=0} \\ \nonumber
    &=
     \left \langle (h_i({\bf x},t)u_k({\bf x},t)- h_k({\bf x},t)u_i({\bf x},t)) \frac{\partial h_j({\bf x},t)}{\partial x_i} \right \rangle \\
     =& \left. 2\frac{\partial A_{[rt]t}^{{\bf u}{\bf h}{\bf h}}(r, t) }{\partial r}\right|_{{r}=0} \delta_{jk}\;.
     \label{eq:ident}
\end{align}
Next, we derive the following identity
\begin{widetext}
\begin{align}\nonumber
 \lefteqn{\left \langle (h_i({\bf x},t)u_k({\bf x},t)- h_k({\bf x},t)u_i({\bf x},t)) \frac{\partial h_j({\bf x},t)}{\partial x_i} \right \rangle}\\ \nonumber
 =&
 \left \langle h_i({\bf x},t)u_k({\bf x},t)\frac{\partial h_j({\bf x},t)}{\partial x_i} \right \rangle - \left \langle h_k({\bf x},t)u_i({\bf x},t)\frac{\partial h_j({\bf x},t)}{\partial x_i} \right \rangle\\ \nonumber
 =& \underbrace{\frac{\partial}{\partial x_i}\left \langle h_i({\bf x},t)h_j({\bf x},t)u_k({\bf x},t)\right \rangle}_{=0,\; \textrm{homogeneity}} -  \left \langle h_i({\bf x},t)h_j({\bf x},t)\frac{\partial u_k({\bf x},t)}{\partial x_i} \right \rangle -
 \left \langle \underbrace{\frac{\partial h_i({\bf x},t)}{\partial x_i}}_{=0}h_j({\bf x},t)u_k({\bf x},t)\right \rangle \\ \nonumber
 &-\underbrace{\frac{\partial}{\partial x_i}\left \langle h_k({\bf x},t)u_i({\bf x},t)h_j({\bf x},t)\right \rangle}_{=0,\; \textrm{homogeneity}} + \left \langle h_j({\bf x},t)u_i({\bf x},t)\frac{\partial h_k({\bf x},t)}{\partial x_i} \right \rangle  +
 \left \langle \underbrace{\frac{\partial u_i({\bf x},t)}{\partial x_i}}_{=0}h_j({\bf x},t)h_k({\bf x},t)\right \rangle  \\
 =&  -  \left \langle h_i({\bf x},t)h_j({\bf x},t)\frac{\partial u_k({\bf x},t)}{\partial x_i} \right \rangle + \left \langle h_j({\bf x},t)u_i({\bf x},t)\frac{\partial h_k({\bf x},t)}{\partial x_i} \right \rangle\;.
 \label{eq:id1}
\end{align}
\end{widetext}
Interchanging $j$ and $k$ in Eq. (\ref{eq:ident}) yields
\begin{align}\nonumber
&\left \langle (h_i({\bf x},t)u_j({\bf x},t) - h_j({\bf x},t)u_i({\bf x},t)) \frac{\partial h_k({\bf x},t)}{\partial x_i} \right \rangle\\ \nonumber
 =& \underbrace{\frac{\partial}{\partial x_i}\left \langle h_i({\bf x},t)h_k({\bf x},t)u_j({\bf x},t)\right \rangle}_{=0,\; \textrm{homogeneity}} \\ \nonumber
 &-  \left \langle h_i({\bf x},t)h_k({\bf x},t)\frac{\partial u_j({\bf x},t)}{\partial x_i} \right \rangle\\ \nonumber
 &-
 \left \langle \underbrace{\frac{\partial h_i({\bf x},t)}{\partial x_i}}_{=0}h_k({\bf x},t)u_j({\bf x},t)\right \rangle \\ 
 &- \left \langle h_j({\bf x},t)u_i({\bf x},t)\frac{\partial h_k({\bf x},t)}{\partial x_i} \right \rangle \;.
 \label{eq:id2}
\end{align}
Adding Eqs. (\ref{eq:id1}) and (\ref{eq:id2}) yields
\begin{align} \nonumber
    &4 \left. \frac{\partial A_{[rt]t}^{{\bf u}{\bf h}{\bf h}}(r, t) }{\partial r}\right|_{{r}=0} \delta_{jk}
    =-  \left \langle h_i({\bf x},t)h_j({\bf x},t)\frac{\partial u_k({\bf x},t)}{\partial x_i} \right \rangle \\
    &- \left \langle h_i({\bf x},t)h_k({\bf x},t)\frac{\partial u_j({\bf x},t)}{\partial x_i} \right \rangle \underbrace{=}_{\textrm{Eq.}(\ref{eq:ident0})}-5\left.\frac{\partial  C_{r\;r\;r}^{\mathbf{h}\mathbf{h} \mathbf{u}}(r,t)}{\partial r}\right|_{r=0} \delta_{jk}\;. 
\end{align}
Hence, for small-scale separations $r$, the defining scalars of symmetric and anti-symmetric mixed third-order tensors are related according to
\begin{equation}
     \left. \frac{\partial A_{[rt]t}^{{\bf u}{\bf h}{\bf h}}(r, t) }{\partial r}\right|_{{r}=0}=-\frac{5}{4}\left.\frac{\partial  C_{r\;r\;r}^{\mathbf{h}\mathbf{h} \mathbf{u}}(r,t)}{\partial r}\right|_{r=0}\;.
\end{equation}
\section{Loitsiansky Invariants and their Implications for the Additional Source Term}
\label{app:recast}
In this section, we discuss further implications of Eqs. (\ref{eq:kh_u}) and (\ref{eq:kh_h}), namely the existence of certain invariants. As shown by Loitsiansky~\cite{loitsianskii1939einige}, in the hydrodynamic limit, the von K\'arm\'an-Howarth equation admits the invariant
\begin{equation}
    \Lambda^{{\bf u}{\bf u}}=\int_0^{\infty} \textrm{d}r\; r^4 C_{r\;r}^{{\bf u}{\bf u}}(r,t)\;.
\end{equation}
As shown by Chandrasekhar~\cite{chandra:1951}, this quantity should also be conserved in MHD turbulence, which can be seen by multiplying Eq. (\ref{eq:kh_u}) by $r^4$ and subsequent integration
\begin{align}\nonumber
   &\frac{\partial}{\partial t}\int_0^r \textrm{d}r' \;r'^4 C_{r\;r}^{\mathbf{u}\mathbf{u}}(r',t)\\
   =&   r^4\left( C_{r\;r\;r}^{\mathbf{u}\mathbf{u} \mathbf{u}}(r,t)-
    C_{r\;r\;r}^{\mathbf{h}\mathbf{h} \mathbf{u}}(r,t) + 2 \nu \frac{\partial}{\partial r} C_{r\;r}^{\mathbf{u}\mathbf{u}}(r,t)\right)\;.
    \label{eq:loit_uu}
\end{align}
Hence, if the bracketed terms on the r.h.s. decay more rapidly than $\sim r^{-4}$ in the limit of $r \rightarrow \infty$, then $\Lambda^{{\bf u}{\bf u}}$ is a conserved quantity in MHD turbulence as well. Nonetheless, due to the anti-symmetry of the third-order correlation in Eq. (\ref{eq:kh_h}), one cannot establish a similar line of reasoning for the magnetic Loitsiansky invariant
\begin{equation}
    \Lambda^{{\bf h}{\bf h}}=\int_0^{\infty} \textrm{d}r\; r^4 C_{r\;r}^{{\bf h}{\bf h}}(r,t)\;.
\end{equation}
Following Chandrasekhar~\cite{chandra:1951}, we introduce the vector potential ${\bf a}({\bf x},t)$ according to
\begin{equation}
{\bf b}({\bf x},t)= \nabla \times {\bf a}({\bf x},t)\;,
\end{equation}
The anti-symmetric tensor (\ref{eq:anti_sym}) can thus be re-cast as
\begin{align} \nonumber
     &A_{[ki]j}^{\mathbf{u}\mathbf{h}{\mathbf{h}}}({\bf r},t)=\langle (u_k({\bf x},t) h_i ({\bf x},t)-u_i({\bf x},t) h_k ({\bf x},t) )h_j({\bf
x}',t)  \rangle\\
=& \varepsilon_{jlm} \frac{\partial}{\partial r_l}\langle (u_k({\bf x},t) h_i ({\bf x},t)-u_i({\bf x},t) h_k ({\bf x},t) )a_j({\bf
x}',t)  \rangle\;.
\label{eq:a_form}
\end{align}
The corresponding tensor is skew and possesses the following form
\begin{align} \nonumber
     &A_{[ki]j}^{\mathbf{u}\mathbf{h}{\mathbf{a}}}({\bf r},t)=\langle (u_k({\bf x},t) h_i ({\bf x},t)-u_i({\bf x},t) h_k ({\bf x},t) )a_j({\bf
x}',t)  \rangle \\
=& 2 A_{[rt]t}^{\mathbf{u}\mathbf{h}{\mathbf{a}}}({r},t) \varepsilon_{ijk} +r\frac{\partial A_{[rt],t}^{\mathbf{u}\mathbf{h}{\mathbf{a}}}({r},t)}{\partial r} \left(\frac{r_i}{r} \varepsilon_{jkl}\frac{r_l}{r} +\frac{r_k}{r} \varepsilon_{jil}\frac{r_l}{r}   \right)\;.
\end{align}
Inserting this tensorial form in Eq. (\ref{eq:a_form}) thus yields
\begin{align}
    A_{[rt]t}^{\mathbf{u}\mathbf{h}{\mathbf{h}}}({r},t)=
    -\frac{1}{r^3} \frac{\partial}{\partial r}\left(r^4 \frac{\partial A_{[rt]t}^{\mathbf{u}\mathbf{h}{\mathbf{a}}}({r},t)}{\partial r}\right)\;,
\end{align}
which upon insertion into Eq. (\ref{eq:kh_h}) yields
\begin{equation}
 \frac{\partial}{\partial t} C_{r\;r}^{\mathbf{h}\mathbf{h}}(r,t) = \frac{1}{r^4} \frac{\partial}{\partial r}\left[r^4 \left(\frac{\partial A_{[rt]t}^{\mathbf{u}\mathbf{h}{\mathbf{a}}}({r},t)}{\partial r} + 2\lambda\frac{\partial}{\partial r} C_{r\;r}^{\mathbf{h}\mathbf{h}}(r,t) \right) \right]\;.
 \end{equation}
Hence, similar arguments as the ones used in Eq. (\ref{eq:loit_uu}) lead to $\lambda^{{\bf h}{\bf h}}= const.$, and Chandrasekhar further concluded $\lambda^{{\bf h}{\bf h}}= 0$. From Eq. (\ref{eq:kh_h}), we can thus derive
\begin{equation}
    \int_{0}^{\infty} \textrm{d}r' r'^3  A_{[rt]t}^{\mathbf{u}\mathbf{h}{\mathbf{h}}}({r},t)= 0\;.
\end{equation}
\section{Longitudinal and Transverse Correlation Functions}
\label{app:chan}
In this Appendix, we derive tensorial forms for second and third-order correlation functions under the assumptions of homogeneity and isotropy. To this end,
we consider the velocity fields ${\bf u}({\bf x}+ {\bf r},t)$ at point ${\bf x}+{\bf r} $ and ${\bf u}({\bf x},t)$ at point ${\bf x}$ (a similar treatment applies for magnetic field). We can divide the vector ${\bf u}= {\bf u}_r +{\bf u}_t$ into a part  ${\bf u}_r$ parallel to ${\bf r}$, and a transverse part ${\bf u}_t$.
These parts are thereby given as
\begin{eqnarray}
 {\bf u}_r= \frac{{\bf r}}{r} \left(\frac{{\bf r}}{r} \cdot {\bf u} \right)\;, \;\; \textrm{and} \;\; 
 {\bf u}_t= -\left(\frac{{\bf r}}{r} \times \left(\frac{{\bf r}}{r} \times {\bf u} \right)\right).
\end{eqnarray}
The longitudinal correlation function
\begin{equation}
 C_{r\,r}(r,t)= \langle {\bf u}_r({\bf x},t) \cdot{\bf u}_r({\bf x}+{\bf r},t) \rangle\;,
 \label{eq:c_longi}
\end{equation}
can be calculated by multiplying the two-point correlation tensor
\begin{equation}
    C_{i\,j}({\bf r},t)=\langle u_i({\bf x},t) u_j({\bf
x}+{\bf r},t)\rangle\;,
\label{eq:C_ij}
\end{equation}
by $\frac{r_i}{r_j}$ and $\frac{r_j}{r}$.
Assuming that $ C_{i\,j}({\bf r},t)=\langle u_i({\bf x},t) u_j({\bf
x}+{\bf r},t)\rangle$ is isotropic and mirror-symmetric~\cite{monin} its general form is given by
\begin{equation}
 C_{i\,j}({\bf r},t)=C_1(r,t) \frac{r_i r_j}{r^2} + C_2(r,t) \delta_{ij}\;,
 \label{eq:c_ij_initial}
 \end{equation}
where the defining scalars $C_1(r,t)$ and $C_2(r,t)$ can now be expressed in terms of the longitudinal (\ref{eq:c_longi})  and transverse correlation functions
\begin{equation}
 C_{t\,t}(r,t)= \frac{1}{2}\langle {\bf u}_t({\bf x}+ {\bf r},t) \cdot {\bf u}_t({\bf x},t)
\rangle\;.
\end{equation}
Here, the factor $\frac{1}{2}$ has been included in three dimensions: two transverse and only one longitudinal direction. Multiplying Eq. (\ref{eq:c_ij_initial}) by $\frac{r_i r_j}{r^2}$ thus yields
\begin{equation}
    C_{rr}(r,t)=C_1(r,t)+C_2(r,t)\;.
    \label{eq:long}
\end{equation}
Decomposing the correlation tensor (\ref{eq:C_ij}) yields
\begin{align}\nonumber
    C_{ij}({\bf r},t) =& \langle {\bf u}_r({\bf x},t) \cdot{\bf u}_r({\bf x}+{\bf r},t) \rangle \frac{r_i r_j}{r^2} \\
    &+ \langle {u}_{t,i}({\bf x}+ {\bf r},t) {u}_{t,j}({\bf x},t)
\rangle\;.
\end{align}
Hence, we can identify
\begin{align} \nonumber
   \lefteqn{\langle {u}_{t,i}({\bf x}+ {\bf r},t) {u}_{t,j}({\bf x},t)
\rangle}\\ \nonumber
=& \underbrace{(C_1(r,t) - C_{rr}(r,t)) }_{\textrm{Eq.} (\ref{eq:long})} \frac{r_i r_j}{r^2}+C_2(r,t) \delta_{ij} \\
=& C_2(r,t) \left(\delta_{ij} - \frac{r_i r_j}{r^2}\right)\;.
\label{eq:c_trans}
\end{align}
Summing Eq. (\ref{eq:c_trans}) over $i=j$ thus yields
\begin{equation}
    2C_{tt}(r,tt) =2C_2(r,t)\;,
\end{equation}
and we can express the correlation tensor (\ref{eq:c_ij_initial}) in terms of its longitudinal and transverse correlation functions according to

\begin{equation}
 C_{i\,j}({\bf r},t)=\left(C_{r\,r}(r,t)-C_{t\,t}(r,t) \right) \frac{r_i r_j}{r^2} + C_{t\,t}(r,t) \delta_{ij}.
 \label{eq:C_{ij}}
 \end{equation}
The bilinear form (\ref{eq:c_ij_initial}) can now be extended to a trilinear form~\cite{Robertson1940}. A general isotropic and mirror-symmetric tensor of order three can be defined as
\begin{align}\nonumber
    C_{ijk}({\bf r},t)=& C_1(r,t) \frac{r_i r_j r_k}{r^3} + C_2(r,t) \frac{r_k}{r} \delta_{ij} \\
    &+C_3(r,t) \frac{r_j}{r} \delta_{ik} +C_4(r,t) \frac{r_i}{r} \delta_{jk}\;.
    \label{eq:trilin}
\end{align}
Whereas a bilinear form is always symmetric in $i$ and $j$, we can now impose further symmetry conditions on this tensor. E.g., the third-order correlation function
\begin{equation}
C_{(ij)k}^{{\bf u}{\bf u}{\bf u}} ({\bf r},t)= \langle u_i({\bf x},t) u_j({\bf x},t) u_k({\bf
x}+{\bf r},t) \rangle\;,
\end{equation}
is symmetric in $i$ and $j$, which implies $C_3(r,t)=C_4(r,t)$. This also applies to the third-order correlation $C_{(ij)k}^{{\bf h}{\bf h}{\bf u}}=\langle h_i({\bf x},t) h_j({\bf x},t) u_k({\bf
x}+{\bf r},t) \rangle$, which stems from the Lorentz force. Both of these tensors can thus be expressed as
\begin{align} \nonumber
 C_{(ij)k}({\bf r},t)&=C_{1}(r,t) \frac{r_i r_j r_k}{r^3} + C_{2}(r,t) \frac{r_k}{r} \delta_{ij}  \\
 &+C_{3}(r,t) \left(\frac{r_j}{r} \delta_{ik} +  \frac{r_i}{r}  \delta_{jk} \right)\;,
 \label{eq:c_ijk}
 \end{align}
 where the coefficients will be specified later on in terms of longitudinal and transverse correlation functions. In contrast, the third-order tensor
 \begin{align}
\lefteqn{A_{[ij]k}^{\mathbf{u}\mathbf{h}{\mathbf{h}}}({\bf r},t)}\\ \nonumber
&=\langle (u_i({\bf x},t) h_j ({\bf x},t)-u_j({\bf x},t) h_i ({\bf x},t) )h_k({\bf
x}+{\bf r},t)  \rangle\;,
\end{align}
is anti-symmetric in $i$ and $j$, which implies $C_1(r,t)=C_2(r,t)=0$ and $C_3(r,t) =- C_4(r,t)= A_1(r,t)$. Hence, an anti-symmetric tensor of order three obeys the trilinear form
\begin{equation}
    A_{[ij]k}({\bf r},t)=A_1(r,t)\left( \frac{r_j}{r} \delta_{ik}- \frac{r_i}{r} \delta_{jk} \right)\;.
\end{equation}
 
\section{The Correlation Functions for Incompressible, Isotropic, and Homogeneous Fields}
\label{app:chan_new}
\subsection{Correlation Functions of Second Order}
Due to the incompressibility condition, it is possible to reduce the tensorial form of $C_{i\,j}({\bf r},t)$ to a dependence of the longitudinal structure function $C_{r\,r}(r,t)$ only. The incompressibility condition is used according to
\begin{equation}
 \frac{\partial}{\partial r_i} C_{i\,j}({\bf r},t)= \left \langle \frac{\partial
u_i({\bf x}+{\bf r},t) }{\partial r_i} u_j({\bf x},t) \right \rangle=0\;.
\label{eq:null}
\end{equation}
Using the relation $\frac{\partial}{\partial r_i}= \frac{r_i}{r} \frac{\partial}{\partial r}$ yields
\begin{eqnarray}
&~&\frac{\partial}{\partial r_i} C_{i\,j}({\bf r},t)
 = \frac{\partial}{\partial r}
(C_{r\,r}(r,t)-C_{t\,t}(r,t)) \frac{r_j}{r} \\ \nonumber
&~&+ \frac{2}{r}
(C_{r\,r}(r,t)-C_{t\,t}(r,t)) \frac{r_j}{r} +
\frac{\partial}{\partial r} C_{t\,t}(r,t) \frac{r_j}{r} =0\;,
\end{eqnarray}
and results in the von K\'arm\'an-Howarth relation
\begin{equation}
  C_{t\,t}(r,t)= \frac{1}{2 r} \frac{\partial}{\partial r} \left( r^2
C_{r\,r}(r,t) \right)\;.
\label{eq:C_tt}
\end{equation}
The correlation function $C_{i\,j}({\bf r},t)$ can therefore be described solely in terms of the longitudinal correlation function $C_{r\,r}(r,t)$. Furthermore, summing $C_{i\,j}({\bf r},t)$ over equal indices $i=j$ we obtain
\begin{align} \nonumber
\sum_{i =j} C_{i\,j}({\bf r},t)=& C_{r\,r}(r,t)+ 2
C_{t\,t}(r,t)\\
=&  \frac{1}{r^2}\frac{\partial}{\partial r} \left( r^3
C_{r\,r} (r,t)  \right)\;.
\end{align}
\subsection{Correlation Functions of Third Order}
Applying the incompressibility condition
\begin{equation}
\frac{\partial}{\partial r_k} C_{(ij)k} ({\bf r},t)=0\;,
\label{eq:inc3}
\end{equation}
to the third-order correlation function (\ref{eq:c_ijk}) yields
\begin{eqnarray} \nonumber
&~&\left(\frac{1}{r^2} \frac{\partial}{\partial r} \left( r^2 C_1(r,t) \right) + 2 r \frac{\partial}{\partial r} \frac{C_3(r,t)}{r} \right) \frac{r_i r_j }{r^2} \\
&~&
+\left( \frac{1}{r^2} \frac{\partial}{\partial r} \left(r^2 C_2(r,t) \right) + 2 \frac{C_3(r,t)}{r} \right) \delta_{ij}=0\;.
\label{eq:inc4}
\end{eqnarray}
Since both brackets in (\ref{eq:inc4}) have to vanish identically to satisfy the equation, we obtain two equations along with (\ref{eq:c_ijk}) for the three pre-factors $C_1(r,t)$, $C_2(r,t)$ and $C_3(r,t)$. This system of equations is solved by
\begin{eqnarray}
  C_1(r,t) &=& -\frac{r^2}{2} \frac{\partial}{\partial r} \left( \frac{  C_{r\,r\,r} (r,t)}{r} \right)\;,\\
 C_2(r,t)&=& -\frac{C_{r\,r\,r} (r,t)}{2},  \\
 C_3(r,t)&=&  \frac{1}{4 r}\frac{\partial}{\partial r} \left ( r^2 C_{r\,r\,r} (r,t) \right)\;.
 \end{eqnarray}
Therefore, the third-order correlation function can be written in terms of $C_{r\,r\,r}(r,t)$ only
\begin{eqnarray}
  &~&C_{(ij)k} ({\bf r},t)= -\frac{r^2}{2} \frac{\partial}{\partial r}
\left( \frac{  C_{r\,r\,r} (r,t)}{r} \right) \frac{r_i r_j
r_k}{r^3}\\ \nonumber
 &+& \frac{1}{4 r}\frac{\partial}{\partial r} \left ( r^2 C_{r\,r\,r}
(r,t)  \right) \left( \frac{r_i}{r} \delta_{jk} 
+ \frac{r_j}{r} \delta_{ik} \right) - \frac{C_{r\,r\,r} (r,t)}{2} \frac{r_k}{r} \delta_{ij}\;.
\label{eq:Cijk}
\end{eqnarray}
\section{Structure Functions of Incompressible, Isotropic, and Homogeneous Fields}
\label{app:struc}
We consider the corresponding moments of velocity increments to calculate the structure functions introduced in Sec.~\ref{sec:derivation}.
\subsection{Structure Functions of Second Order}
The second-order longitudinal structure functions are defined as
\begin{align}
    S_{r\;r}^{{\bf u}{\bf u}}(r,t) =& \left \langle \left( \left[{\bf u}({\bf x}+{\bf r},t)- {\bf u}({\bf x},t) \right]\cdot \frac{{\bf r}}{r} \right)^2\right  \rangle\;,  \\
     S_{r\;r}^{{\bf h}{\bf h}}(r,t) =& \left \langle \left( \left[{\bf h}({\bf x}+{\bf r},t)- {\bf h}({\bf x},t) \right]\cdot \frac{{\bf r}}{r} \right)^2\right  \rangle\;.  
\end{align}
These longitudinal structure functions can be related to the longitudinal correlation functions, which will be shown here based on the longitudinal velocity correlation function $C_{r\;r}^{{\bf u}{\bf u}}(r,t)$ defined by Eq.(\ref{eq:c_longi}). We obtain
\begin{align}
    \lefteqn{\left \langle \left( \left[{\bf u}({\bf x}+{\bf r},t)- {\bf u}({\bf x},t) \right]\cdot \frac{{\bf r}}{r} \right)^2\right  \rangle}\\ \nonumber
    =& \frac{r_i}{r} \left \langle {u}_i({\bf x}+{\bf r},t) {u}_j({\bf x}+{\bf r},t) + {u}_i({\bf x},t) {u}_j({\bf x},t) \right \rangle  \frac{r_j}{r} \\ \nonumber
    &-\frac{r_i}{r} \left \langle {u}_i({\bf x}+{\bf r},t) {u}_j({\bf x},t)-{u}_i({\bf x},t) {u}_j({\bf x}+{\bf r},t) \right \rangle  \frac{r_j}{r}. 
\end{align}
Due to homogeneity and isotropy, we obtain
\begin{align}\nonumber
   &\left \langle u_i({\bf x},t)  u_j({\bf x},t) \right \rangle= \left \langle u_i({\bf x}+{\bf r},t)  u_j({\bf x}+{\bf r},t) \right \rangle= C_{i\;j}^{{\bf u}{\bf u}}({\bf 0},t)\\ \nonumber
   &\left \langle u_i({\bf x},t)  u_j({\bf x}+{\bf r},t) \right \rangle= \left \langle u_i({\bf x}+{\bf r},t)  u_j({\bf x},t) \right \rangle= C_{i\;j}^{{\bf u}{\bf u}}({\bf r},t)\;.
\end{align}
Hence, we obtain 
\begin{align}
    S_{r\;r}^{{\bf u}{\bf u}}(r,t)=&  \frac{r_i}{r} \left[2C_{i\;j}^{{\bf u}{\bf u}}({\bf 0},t)-2 C_{i\;j}^{{\bf u}{\bf u}}({\bf r},t)\right]\frac{r_j}{r} \\ 
    =& 2 \left[C_{r\;r}^{{\bf u}{\bf u}}({ 0},t)- C_{r\;r}^{{\bf u}{\bf u}}({r},t)\right]\;.
\end{align}
The same treatment applies to the magnetic structure function of the second order.
\subsection{Structure Functions of Third Order}
The third-order longitudinal velocity structure function
\begin{equation}
    S^{{\bf u}{\bf u}{\bf u}}_{r\;r\;r}(r,t) =\left \langle \left( \left[{\bf u}({\bf x}+{\bf r},t)- {\bf u}({\bf x},t) \right]\cdot \frac{{\bf r}}{r} \right)^3\right \rangle
\end{equation}
can be related to the third-order longitudinal correlation function as
\begin{align}\nonumber
    \lefteqn{S^{{\bf u}{\bf u}{\bf u}}_{r\;r\;r}(r,t) }\\ \nonumber
    =&\left \langle \left[{u}_i({\bf x}+{\bf r},t)- {u}_i({\bf x},t) \right] 
    \left[{u}_j({\bf x}+{\bf r},t)- {u}_j({\bf x},t) \right] \right. \\ \nonumber
    ~&\left.\left[{u}_k({\bf x}+{\bf r},t)- {u}_k({\bf x},t) \right]\right \rangle \frac{r_i r_j r_k}{r^3}\\ \nonumber
    =& 2 \left[C_{(ij)k}^{{\bf u}{\bf u}{\bf u}} ({\bf r},t) + C_{(jk)i}^{{\bf u}{\bf u}{\bf u}} ({\bf r},t)+ C_{(ik)j}^{{\bf u}{\bf u}{\bf u}} ({\bf r},t) \right]\frac{r_i r_j r_k}{r^3}\\
    =& 6 C_{r\;r\;r}^{{\bf u}{\bf u}{\bf u}}(r,t)\;. 
\end{align}
Here, we used statistical homogeneity and isotropy and the relation $\left \langle u_i({\bf x},t)u_j({\bf x},t)u_k({\bf x},t) \right \rangle=0$.

\bibliographystyle{apsrev4-1}
\bibliography{mhd_long_trans.bib}
\end{document}